\documentclass[useAMS,usenatbib]{mn2e}

\usepackage{natbib} 
\usepackage{amsmath} 
\usepackage{amsbsy}
\usepackage{amssymb}
\usepackage{mathrsfs} 
\usepackage{lscape} 
\usepackage{graphicx}
\usepackage{epstopdf}
\usepackage{deluxetable} 
\usepackage{fixltx2e} 
\usepackage{subfig}
\usepackage{hyperref}
\usepackage{verbatim}
\usepackage[inline]{enumitem}

\setlist[enumerate]{leftmargin=*}

\usepackage{color,soul}

\newcommand{\NHI}{N_{\textrm{HI}}}
\newcommand{\Mvir}{M_\textrm{h}}
\newcommand{\Rvir}{R_\textrm{vir}}
\newcommand{\Msun}{{\rm M}$_\odot$}

\newcommand{\logMvir}{\log{\Mvir}}

\newcommand{\matr}[1]{\mathbf{#1}}

\title[Low-$z$ Lyman Limit Systems in the FIRE Simulations]{Low-Redshift Lyman Limit Systems as Diagnostics of Cosmological Inflows and Outflows}

\author[Hafen et al.]{
\parbox{\textwidth}{
Zachary Hafen,$^{1}$\thanks{zhafen@u.northwestern.edu}
Claude-Andr\'e Faucher-Gigu\`ere,$^{1}$
Daniel Angl\'es-Alc\'azar,$^1$
Du\v{s}an Kere\v{s}$,^2$
Robert Feldmann,$^{3,4}$
T. K. Chan,$^2$
Eliot Quataert,$^3$
Norman Murray,$^5$
Philip F. Hopkins$^6$ 
} \vspace{0.4cm}\\
\parbox{\textwidth}{
$^{1}$Department of Physics and Astronomy and Center for Interdisciplinary Exploration and Research in Astrophysics (CIERA),\\ Northwestern University, 2145 Sheridan Road, Evanston, IL 60208, USA. \\
$^2$Department of Physics, Center for Astrophysics and Space Sciences, University of California, San Diego, 9500 Gilman Drive, \\ La Jolla, CA 9209, USA \\
$^3$Department of Astronomy and Theoretical Astrophysics Center, University of California, Berkeley, CA 94720-3411, USA \\
$^4$Institute for Computational Science, University of Zurich, Zurich CH-8057, Switzerland \\
$^5$Canadian Institute for Theoretical Astrophysics, 60 St George Street, University of Toronto, ON M5S 3H8, Canada \\
$^{6}$TAPIR, Mailcode 350-17, California Institute of Technology, Pasadena, CA 91125, USA \\
}
}

\begin{document}
\maketitle

\begin{abstract}
We use cosmological hydrodynamic simulations with stellar feedback from the FIRE project to study the physical nature of Lyman limit systems (LLSs) at $z\le1$. 
At these low redshifts, LLSs are closely associated with dense gas structures surrounding galaxies, such as galactic winds, dwarf satellites, and cool inflows from the intergalactic medium. 
Our analysis is based on 14 zoom-in simulations covering the halo mass range $M_{\rm h}\approx10^{9}-10^{13}$ M$_{\odot}$ at $z=0$, which we convolve with the dark matter halo mass function to produce cosmological statistics.
We find that the majority of cosmologically-selected LLSs are associated with halos in the mass range $10^{10} \lesssim \Mvir \lesssim 10^{12}$ M$_{\odot}$. 
The incidence and HI column density distribution of simulated absorbers with columns $10^{16.2} \leq N_{\rm HI} \leq 2\times10^{20}$ cm$^{-2}$ are consistent with observations. 
High-velocity outflows (with radial velocity exceeding the halo circular velocity by a factor $\gtrsim2$) tend to have higher metallicities ($[X/H]\sim -0.5$) while very low metallicity ($[X/H]<-2$) LLSs are typically associated with gas infalling from the intergalactic medium. 
However, most LLSs occupy an intermediate region in metallicity-radial velocity space, for which there is no clear trend between metallicity and radial kinematics. 
The overall simulated LLS metallicity distribution has a mean (standard deviation) [X/H] $=-0.9$ (0.4) and does not show significant evidence for bimodality, in contrast to recent observational studies but consistent with LLSs arising from halos with a broad range of masses and metallicities. 
\end{abstract}

\begin{keywords}
galaxies: formation --- galaxies: evolution --- galaxies: haloes --- quasars: absorption lines --- intergalactic medium --- cosmology: theory\vspace{-0.5cm}
\end{keywords}

\section{Introduction}

The circum-galactic medium (CGM) plays a vital role in determining the evolution of galaxies. Accretion from the intergalactic medium (IGM) passing through the CGM provides the fuel to sustain star formation in galaxies over a Hubble time \citep[e.g.][]{Keres2005, Prochaska2009, Bauermeister2010}. At the same time, galactic feedback processes driven by stars and active galactic nuclei are observed to expel gas out of galaxies  and into the CGM as outflows  \citep[e.g.,][]{Heckman2000, Shapley2003, Martin2005, Weiner2009, Steidel2010, Greene2011, Jones2012, Rubin2014, Cicone2014}. 
Cosmological models of galaxy formation furthermore show that strong outflows are necessary to regulate galaxy growth \citep[e.g.,][]{Keres2009, AnglesAlcazar2014, Hopkins2014, Somerville2015} and explain the metal enrichment of the IGM \citep[][]{Aguirre2001, Oppenheimer2006, Booth2012}. 
At low redshift, Lyman limit systems (LLSs; loosely defined as systems optically thick at the Lyman limit) trace overdense structures in the halos of galaxies and are thus useful observational probes of inflows and outflows \citep[e.g.,][]{Ribaudo2011a, Tripp2011}. 

In the past few years, the observational census of $z\leq1$ LLSs has been greatly improved by new analyses of \textit{Hubble Space Telescope} (HST) observations. 
\cite{Ribaudo2011} measured the incidence of LLS and column density distribution of HI absorbers, using archival data from the Faint Object Spectrograph (FOS) and Space Telescope Imaging Spectrograph (STIS) instruments. 
Thanks to the installation of the Cosmic Origins Spectrograph (COS) on HST, the increased sample of low-redshift HI-selected LLSs makes it now possible to measure the unbiased metallicity distribution of LLSs. 
\cite{Lehner2013} measured this metallicity distribution and found evidence for a metallicity bimodality.
This measurement included all systems with neutral hydrogen column density $16.2 \leq \log{N_{\rm HI}} \leq 18.5$. 
Here and in the rest of this paper, logarithms are to the base 10, $N_{\rm HI}$ is in units of cm$^{-2}$, and $M_{\rm h}$ is in units of $M_{\odot}$.  
\cite{Lehner2013} tentatively identified the metal-rich branch as likely tracing galactic winds, recycled outflows, and gas tidally-stripped from galaxies. 
They also noted that the metal-poor branch has properties consistent with the cold accretion streams predicted by cosmological simulations \citep[e.g.,][]{Keres2005, Keres2009, Dekel2006, Dekel2009, Brooks2009, vdVoort2011, FaucherGiguere2011}. 
In an analysis that doubles the Lehner et al. absorber sample, \cite{2016ApJ...831...95W} find broadly consistent evidence for a metallicity bimodality, but note that the bimodality primarily arises from the (more numerous) partial LLSs with column in the range $16.2 \leq \log{N_{\rm HI}} \leq 17.2$. 
\cite{2016MNRAS.458.4074Q} reported hints of a bimodality in the metallicity distribution of super LLSs (SLLSs) with $19 \leq \log{N_{\rm HI}} \leq 20.3$ (which they refer to as sub damped Ly$\alpha$ absorbers) at $z<1.25$, but with lower statistical significance and with no evidence of a metallicity bimodality at higher redshift.
 \cite{Fox2013} found evidence that metal-rich low-$z$ LLSs tend to show higher O VI columns and broader O VI profiles than metal-poor LLSs. 
 
At the same time, the incidence and metallicity properties of high-redshift LLSs have become increasingly well constrained by ground-based observations \citep[][]{Fumagalli2011b, Fumagalli2013b, Fumagalli2015, Lehner2014, Cooper2015, Prochaska2015}. 
Despite this tremendous observational progress, forward modeling of IGM absorbers using cosmological simulations remains critical to understanding the physical nature of LLSs, since spectroscopy only provides direct information on the line-of-sight structure of absorbers.  

Overall, cosmological simulations have proved very successful at reproducing the observed column density distribution of HI absorbers \citep[][]{Hernquist1996, McQuinn2011, Altay2011, Altay2013, Rahmati2013, 2016arXiv160803293G}, demonstrating how this technique can be used to inform the physical nature of absorption systems. 
In general, the agreement between cosmological simulations and observations is best for the Lyman-$\alpha$ forest \citep[e.g.,][]{Bird2013}, which is relatively little affected by feedback processes from galaxies \citep[e.g.,][]{Kollmeier2006}. 
On the other hand, simulations predict that a large fraction of circum-galactic LLSs trace galactic winds \citep[][]{Fumagalli2014, Rahmati2015a, Faucher-Giguere2015, Liang2015, Faucher-Giguere2016} and the distribution of LLSs is therefore sensitive to feedback processes. 
Most of the other circum-galactic LLSs arise in streams of infalling gas \citep[e.g.,][]{Faucher-Giguere2011, Fumagalli2011, vdVoort12, Goerdt2012, Faucher-Giguere2015} or on the outskirts of galaxies. 
To date, however, most simulation analyses have focused on the properties of LLSs at high redshift \citep[see also][]{Kohler2007}.

Our goal in this paper is to study the properties of simulated LLSs at $z \leq 1$, with an eye toward interpreting recent HST measurements and developing inflow-outflow diagnostics. 
We use a set of cosmological zoom-in simulations from the FIRE project (``Feedback In Realistic Environments'')\footnote{Project website: http://fire.northwestern.edu}, which implement models for stellar feedback on the scale of star-forming regions. 
These simulations have been shown to correctly reproduce the star formation histories of galaxies below $\sim L^{*}$ \citep[][]{Hopkins2014}, mass-metallicity relations \citep[][]{Ma2016} at all redshifts where observations are available, and LLS covering fractions in the halos of $z\sim2$ galaxies \citep[][]{Faucher-Giguere2015, Faucher-Giguere2016}. 
In these simulations, galaxy-scale outflows are self-consistently generated from the local injection of feedback momentum and energy on small scales \citep{Muratov2015, 2016arXiv160609252M}. 
These successes of the FIRE simulations make them particularly well suited to address the physical nature of LLSs. 

For this paper, we define partial LLSs as systems with $16.2 \leq \log{N_{\rm HI}} \leq 17.2$, LLSs as systems with $17.2 \leq \log{N_{\rm HI}} \leq 19$, and SLLSs as systems with $19 \leq \log{N_{\rm HI}} \leq 20.3$. 
Since partial LLSs and LLSs trace similar structures in the CGM, we will often group them together and refer to them simply as LLSs. 
Throughout, we assume a standard flat $\Lambda$CDM cosmology with $h \approx 0.7$, $\Omega_{\rm{M}}\approx0.27$, and $\Omega_{\rm{b}}\approx0.046$, consistent with the latest observational constraints \citep[e.g.,][]{Planck2015}.

The structure of this paper is as follows. 
In \S\ref{methodology}, we describe in more detail the simulations used in our analysis and our methodology to extract observational statistics.  In section \ref{abs_HI_stats}, we convolve our zoom-in simulations results with the dark matter halo mass function to predict cosmological statistics, including the incidence of LLSs and the HI column density distribution of CGM absorbers. 
We quantify relationships between LLS column density, metallicity, and radial velocity relative to central galaxies in \S \ref{hist_section}. 
In that section, we also compute the overall metallicity distribution of low-redshift LLSs predicted by our simulations. 
We discuss our results and conclude in \S\ref{conclusion}. 
The Appendix summarizes convergence tests done to validate our analysis.

\section{Methodology}
\label{methodology}

\subsection{Simulation sample}
Our analysis is based on cosmological zoom-in \citep[e.g.,][]{Porter1985, Katz1993} simulations from the FIRE project. 
As in previous FIRE papers, all SPH simulations presented in this paper were run with exactly the same code as the original FIRE simulations presented in \cite{Hopkins2014}. Therefore, we only summarize the key points here. 
Our CGM analysis methods are similar to the ones used in \cite{Faucher-Giguere2015} and more details can be found in that paper. 

Our simulations were run with the \textsc{P-SPH} \citep{Hopkins2013d} hydrodynamics solver implemented in the \textsc{GIZMO} code \citep[][]{Hopkins2015}. 
P-SPH uses a pressure-entropy formulation of the smooth particle hydrodynamics (SPH) equations that eliminates the artificial surface tension at contact discontinuities found in traditional density-based SPH formulations \citep[e.g.,][]{Agertz2007} and resolves the major historical differences between SPH and grid-based codes. 
The gravity solver in GIZMO is a modified version of the {\small GADGET-3} gravity algorithm \citep{Springel2005} which implements the adaptive softening method of \cite{Price2007} (which we use for the gas) and a modified softening kernel that represents the exact solution for the potential of the SPH smoothing kernel following \cite{Barnes2012}. 

In our simulations, gas is allowed to cool to $\sim10$ K via atomic and molecular line emission, in addition to the standard processes described by \cite{Katz1996}. 
The FIRE simulations are designed to resolve the Toomre mass/length of gas within galaxies, and therefore the most massive giant molecular clouds (GMCs) in which most star formation occurs in typical galaxies \citep[e.g.,][]{Murray2011}. 
In the simulations, star formation occurs only in dense, locally self-gravitating, and molecular/self-shielding gas. 
Where all these criteria are met, the gas is converted into stars on a local free-fall time. 
The density threshold for star formation is set $n_{\rm H} > 50$ cm$^{-3}$ in most of our simulations, though in practice the self-gravity criterion is generally the most stringent. 
Stellar feedback is modeled by implementing energy, momentum, mass, and metal return from radiation pressure, photoionization, photoelectric heating, supernovae, and stellar winds following the STARBURST99 stellar population synthesis model \citep[][]{Leitherer1999}. 
We explicitly follow the chemical abundances of nine metal species (C, N, O, Ne, Mg, Si, S, Ca, and Fe).
During the hydrodynamic simulations, ionization balance of all elements is computed using the cosmic ultraviolet (UV) background model of \cite{Faucher-Giguere2009} and we apply an on-the-fly correction for dense, self-shielded gas based on a local Jeans-length approximation.

\begin{table*}
\caption{Parameters of the New Simulations}
\centering
\begin{tabular}{cccccccc}
\hline
Name & $\Mvir(z=0)$     & $M_*(z=0)$     & $R_{\textrm{vir}}(z=0)$ & $m_\textrm{b}$       & $\epsilon_\textrm{b}$ & $m_{\textrm{dm}}$ & $\epsilon_{\textrm{dm}}$ \\
     & (\Msun) & (\Msun) & (kpc)                & (\Msun) & (pc)         & (\Msun)        & (pc)                      \\
     \hline
m11.4a & 2.6e11      & 6.2e9       & 180                  & 3.3e4       & 9          & 1.7e5              &   140                        \\
m11.9a & 8.4e11      & 3.0e10      & 250                  & 3.4e4       & 9           & 1.7e5              &   140                        \\
MFz0\_A2 & 1.0e13      & 2.7e11      & 630                  & 3.0e5       & 9           & 1.4e6              &      140           \\    
\hline     
\end{tabular}
\\

$\Mvir(z=0)$ and $M_*(z=0)$ are the dark matter and stellar masses inside the present-day virial radius, $R_{\textrm{vir}}(z=0)$, defined  as in \cite{Bryan1998}. 
$m_\textrm{b}$ and $m_{\textrm{dm}}$ are the baryonic and dark matter particle masses. The simulations use adaptive gravitational softening lengths for the gas but fixed softening lengths for the dark matter. 
$\epsilon_\textrm{b}$ is the minimum force softening length for the gas and $\epsilon_{\textrm{dm}}$ is Plummer-equivalent gravitational softening length for the dark matter.
\label{simulation_table}
\end{table*}

We analyze a sample of 14 simulations run to $z=0$ and spanning the halo mass range $M_{\rm h}(z=0) \sim 10^{9} \-- 10^{13} M_\odot$. From previous work, our sample includes the seven main simulations in~\cite{Hopkins2014} (m09, m10, m11, m12v, m12q, m12i, and m13) and four $M_{\rm h}(z=0) \sim 10^{10} \-- 10^{11} M_\odot$ simulations from \cite{Chan2015} (m10h1297, m10h1146, m10h573, and m11h383). 
We ran three new simulations (m11.4a, m11.9a, MFz0\_A2) with $M_{\rm h}(z=0) \sim 10^{11} \-- 10^{13} M_\odot$ to better populate that halo mass range for our analysis. 
Simulation MFz0\_A2 follows the same main halo as the MFz2\_A2 MassiveFIRE simulation \citep[][]{2016arXiv161002411F}, but include a larger high-resolution region and is evolved to $z=0$. 
The simulation parameters for the three new runs are detailed in Table~\ref{simulation_table}. With the exception of m13 and MFz0\_A2, all simulations in our sample have a gas particle mass $m_\textrm{b} \le 5\times10^{4}$~\Msun, minimum adaptive baryonic smoothing/force softening length $\epsilon_\textrm{b} \le 10~\textrm{proper~pc}$, dark matter particle mass $m_{\textrm{dm}} \le 5 \times 10^5$~\Msun, and dark matter force softening length $\epsilon_{\textrm{dm}} = 150~\textrm{proper~pc}$. 
For dwarf galaxies in the sample, the resolution parameters are significantly better. 
For example, the m09 and m10 runs have a gas particle mass $m_{\rm b} \approx 260$ M$_{\odot}$ and correspondingly smaller minimum softening lengths.

\subsection{Extraction of absorber statistics} \label{extraction}
\begin{figure*}
\centering
\begin{minipage}{0.495\textwidth}
\centering
\includegraphics[width=\textwidth, keepaspectratio]{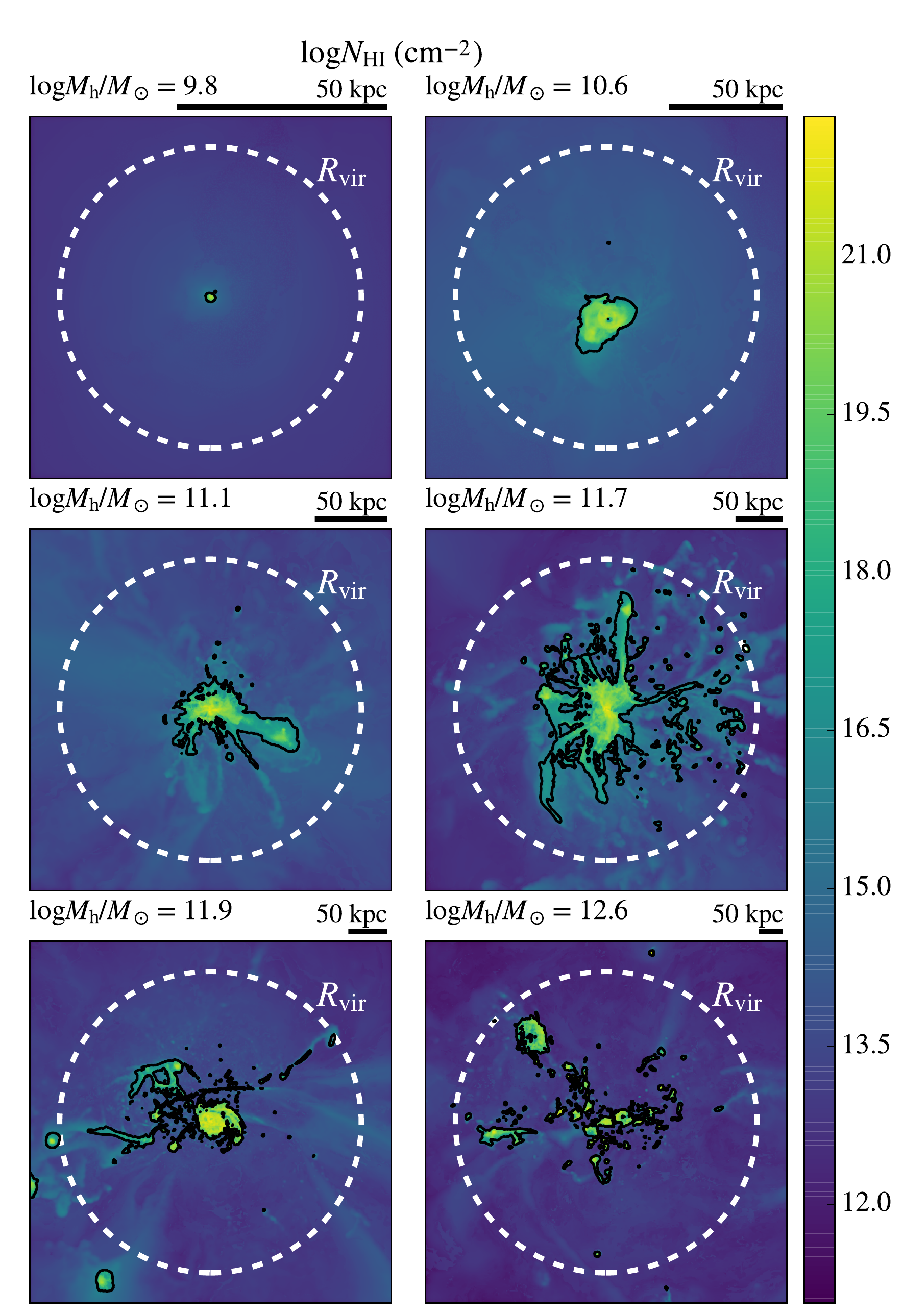}
\end{minipage} \hfill
\begin{minipage}{0.495\textwidth}
\centering
\includegraphics[width=\textwidth, keepaspectratio]{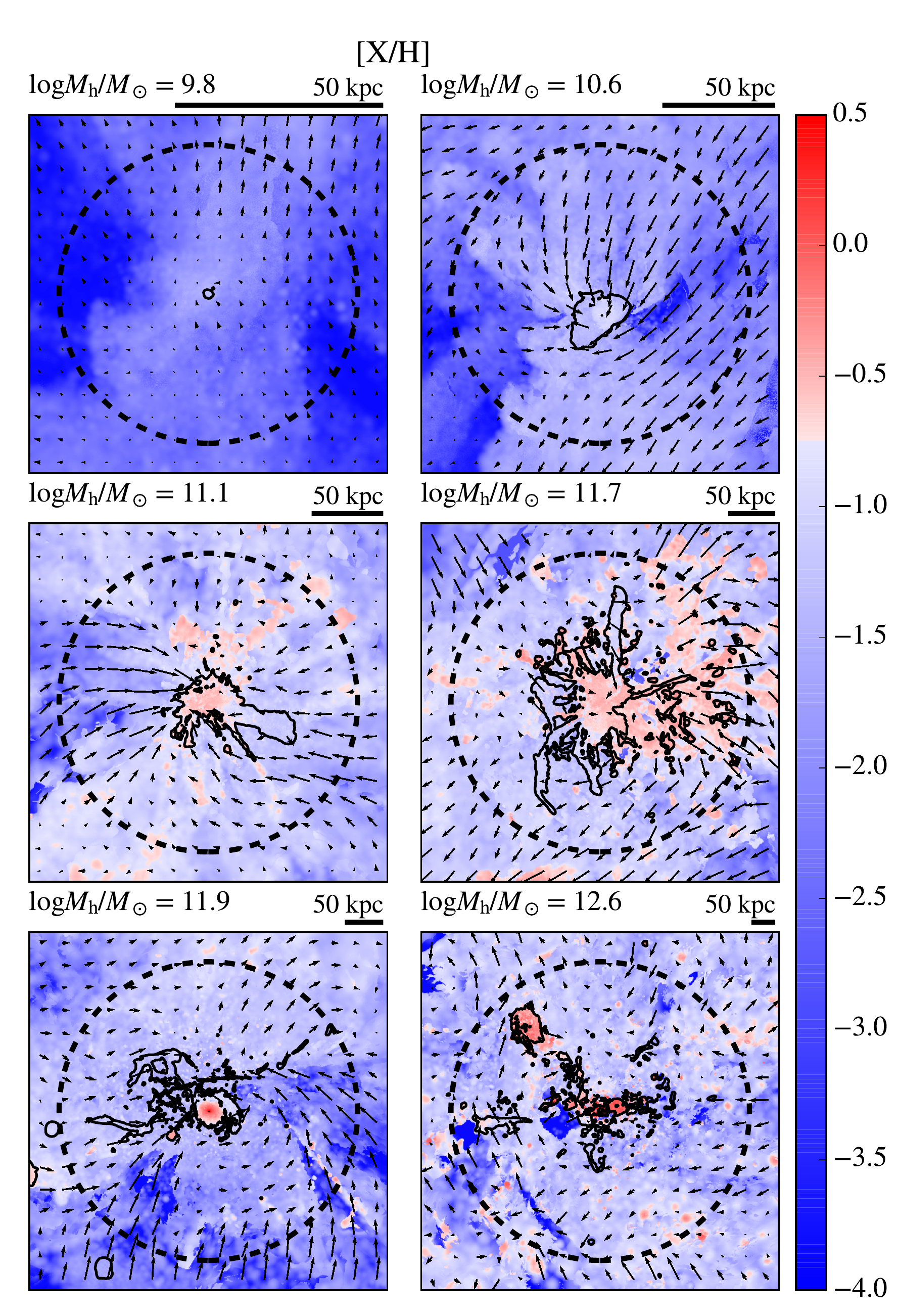}
\end{minipage}
\caption{\textbf{Left:} HI column density maps at $z = 0.5$ for simulations m10, m10h573, m11h383, m11.9a, m12i, and m13 in reading order from top left to bottom right.
Black contours indicate $N_{\rm{H I}} \ge 10^{16.2}$ cm$^{-2}$ (partial Lyman limit systems and higher). 
\textbf{Right:} Same as left, but metallicity in solar units for the same halos as on the left. 
The arrows indicate projected velocity vectors in units of the halo circular velocity.
The HI column density and metallicity shown in these maps are for the strongest absorber along each sight line, calculated as described in \S\ref{extraction}.
Overall, the circum-galactic gas dynamics and their relationship to gas metallicity are complex.
\label{snapshot_summary}
}
\end{figure*}

For each simulation, we center our analysis on the most massive main halo within the zoom-in region, and include 20 snapshots evenly spaced in redshift over $0\le z<0.5$ and 11 evenly spaced snapshots over $0.5\le z \le 1$. 
We use Amiga Halo Finder (AHF) \citep{Knollmann2009} and the virial overdensity definition from \cite{Bryan1998} to identify halos. 
We denote the virial radius $R_{\rm{vir}}$. 
Figure \ref{snapshot_summary} shows HI and metallicity maps for six of our simulated halos at $z=0.5$. 
From the panels on the right, which show velocity vectors overlaid on top of the metallicity maps, we anticipate a complex relationship between instantaneous velocity and metallicity in the CGM of our simulated halos at this redshift, a point to which we will return to quantitatively in \S \ref{hist_section}.

To extract the properties of individual CGM absorbers, we first grid the SPH particle data onto cubic 3D meshes, each with $512^3$ cells. 
The full SPH kernel is used to interpolate particle data onto the grid. 
In cases where the grid does not resolve individual particles, the cloud-in-cell method is used to distribute the particle mass to the nearest 8 grid points in a mass-conserving manner. 
We verified the convergence of our results with respect to grid resolution (see Appendix~\ref{convergence_properties}).
At $z\leq1$, our simulations indicate that the vast majority of LLSs occur within the virial radius of halos. 
To capture this halo cross section, we use grids of side length $2.4~R_{\rm{vir}}$ for each main halo. 
In any given snapshot, we find that up to $\sim20\%$ of LLSs in our grids are found $>R_{\rm vir}$ away from the central galaxy, but almost all are associated with nearby galaxies that have their own main halos. Thus, such LLSs are accounted for by our convolutions over the dark matter halo mass function described in \S \ref{abs_HI_stats}. 
The HI density for each cell is computed using a self-shielding-corrected photoionization rate \citep{Rahmati2013}, which is a function of the total hydrogen density and the ultraviolet background. 
We do this by post-processing our simulated grids, but have verified that computing the HI density on the particles and then depositing that density into cells gives similar results.

\begin{figure}
\begin{center}
\includegraphics[width=\linewidth, keepaspectratio]{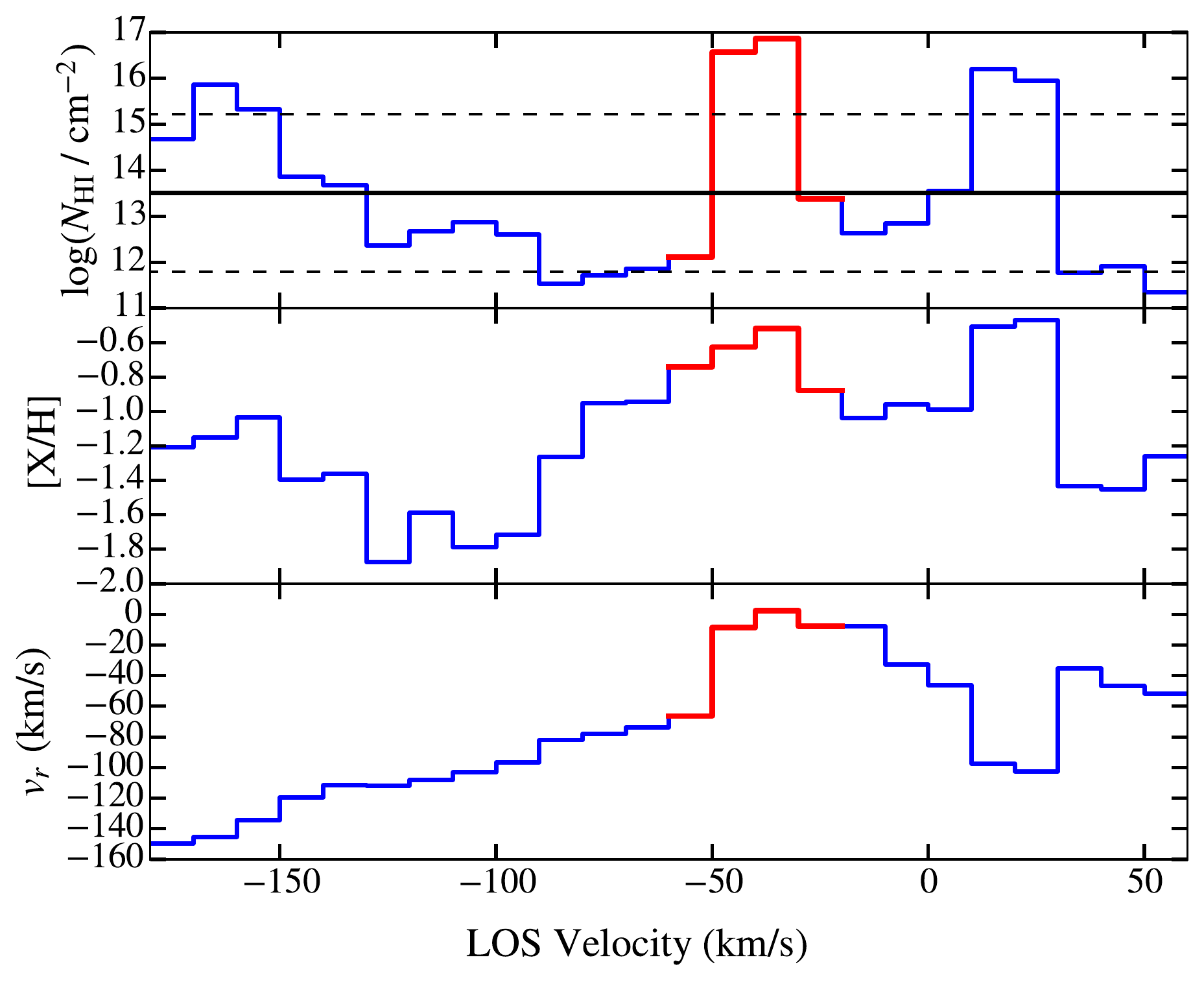}
\caption{Example sight line profiles, as a function of line-of-sight velocity, used to identify and quantify the properties of CGM absorbers. This is a random sight line through the main halo of our m11.9a simulation, with zero LOS velocity corresponding to the galaxy velocity.
\textbf{Top:} $\log$ of HI column density. The horizontal solid black line indicates the average value of $\log{N_{\rm HI}}$ over the sight line, and the dashed lines show $\pm 1 \sigma$ intervals.
\textbf{Middle:} Metallicity profile in solar units.
\textbf{Bottom:} Gas mass-weighted radial velocity relative to the central galaxy.
For each sight line, we select the strongest HI absorber (highlighted in red) using the shape of the $\log{N_{\rm HI}}$ profile and extract the corresponding metallicity and radial velocity.
}
\label{indiv_LOS_prof}
\end{center}
\end{figure}

For each line of sight (LOS), we extract the properties of the strongest HI absorber identified in velocity space.
To select the strongest HI component along each LOS, we use the LOS $\log{N_{\rm HI}}$ profile binned as a function of LOS velocity. 
The strongest component is identified by the peak ${N_{\rm HI}}$ along the LOS, and its width is set by the points at which $\log{N_{\rm HI}}$ drops below its mean value along the LOS. 
We experimented with different quantitative criteria for automatically identifying strongest absorbers along each LOS but found that there is in general no ambiguity in our analysis. This is because there is typically a single strongest absorber along each LOS, well separated in velocity space from other column density peaks.
This approach is a good proxy for how LLSs are identified in observations \citep[e.g., ][]{Lehner2013} and is more accurate than, for example, estimating absorber metallicities by projecting the total gas and metal masses across the simulation volume.
The latter approach can be significantly biased when the metallicity is not uniform along the LOS, which is often the case in the multiphase CGM.
We do not analyze the properties of weaker absorbers.
Figure~\ref{indiv_LOS_prof} illustrates this procedure for a random LOS through the m11.9a halo at $z=0.5$. 

For each absorber identified in this way, we evaluate the HI column density $N_{\rm HI}$, metallicity [X/H], 
and the gas mass-weighted radial velocity $v_{\rm r}$ of the component relative to the central galaxy. We define the metallicity [X/H] of an absorber as $\log (Z/Z_\odot)$, where $Z$ is the metal mass fraction of the absorber gas and $Z_\odot=0.014$ is the solar metal mass fraction \citep{Asplund2010}.

\section{Cosmological H~I Statistics}
\label{abs_HI_stats}
Our zoom-in simulations provide us with predictions for the distribution of absorbers within the virial radius of galaxies as a function of halo mass and redshift. 
Observations that randomly select quasars on the sky, on the other hand, probe a cosmological distribution of absorbers that receives contributions from halos of all masses. 
In this section, we convolve our zoom-in results with the dark matter halo mass function to derive predictions for the cosmological incidence and column density distribution of HI CGM absorbers.

\subsection{Weighting halos as a function of mass and redshift}
\label{diff_mass_dist}

\begin{figure}
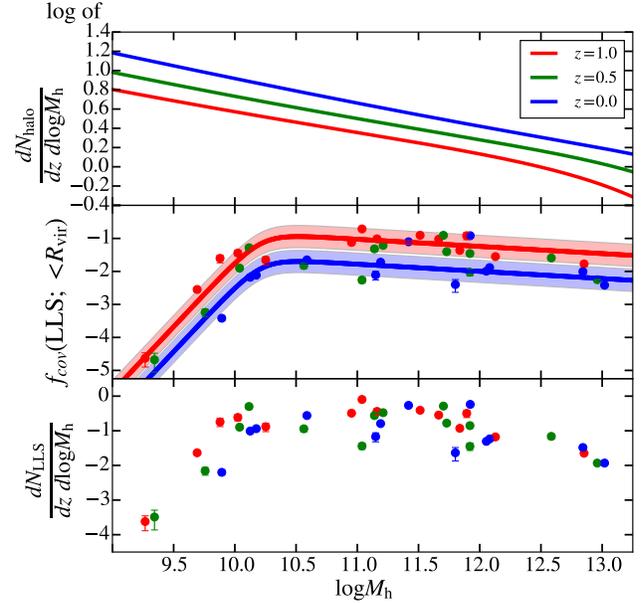

\begin{center}
\includegraphics[width=\linewidth, keepaspectratio]{{{threepanel}}}
\caption{\textbf{Top:} Incidence of dark matter halos (virial cross sections per sight line) per unit redshift and halo mass bin. 
\textbf{Middle:} Covering fraction of LLSs ($16.2 \leq \log{N_{\rm HI}} \leq 19$) within a projected virial radius as a function of halo mass, as predicted by our zoom-in simulations.  A redshift-dependent fit to the covering fractions is shown at $z=0$ and $z=1$ in solid blue and red respectively, with the shaded regions showing the standard deviation of the fit. 
\textbf{Bottom:} Incidence of LLS per unit redshift per unit $\logMvir$. 
The majority of randomly selected LLS are associated with halos in the mass range $10^{10} M_\odot \lesssim \Mvir \lesssim 10^{12} M_\odot$. 
In each panel, the solid dots correspond to the covering fractions averaged over three orthogonal sky projections and the error bars (usually smaller than the solid dot) show the sample standard deviation among the different projections.
}
\label{three_panel}
\end{center}
\end{figure}

Consider a prescribed $N_{\rm HI}$ range. We define $dN_{\rm abs}/dz$ to be the mean number of absorbers in this range per sight line per unit redshift. 
For a given redshift $z$, we quantify the contribution of halos of different mass bins to the total $dN_{\rm abs}/dz$ using the weights
\begin{equation}
w_{\rm{abs}}(z, \Mvir) \equiv \frac{d^2N_{\rm{abs}}}{dz~d\logMvir}.
\end{equation} 
We evaluate $w_{\rm{abs}}(z, \Mvir)$ as follows. 

For any main halo, we define the halo cross section as $\pi R_{\rm vir}^2$. The standard dark matter halo mass function provides the mean number density of halos per halo mass bin as a function of redshift, $dn_{\rm h}/d\log{M_{\rm h}}$. 
The three-dimensional halo number density can be converted into a surface density per unit redshift using the cosmological line element, $dl/dz$. 
We can then simply evaluate the mean number of halos intersected per unit redshift per halo mass bin for random sight lines:
\begin{equation}
\frac{d^2N_{\rm{h}}}{dz~d\logMvir} = \pi R_{\rm vir}^2\frac{dl}{dz}\frac{dn_{\rm h}}{d\logMvir}.
\label{dN_halo}
\end{equation}
The top panel of Figure~\ref{three_panel} shows this quantity at $z = 0,~0.5,$ and 1. For the halo mass function, we used the CPMSO+AHF fit to $N$-body cosmological simulations from \cite{Watson2013a}. 
As expected, the curve increases with decreasing redshift (as halos grow) and decreases with increasing halo mass (as halos become rarer).

To predict the contribution of different halos to the cosmological incidence of absorbers, we need to know the fraction of the halo cross section $\pi R_{\rm vir}^{2}$ covered by the absorbers of interest. 
We define this quantity as the covering fraction $f_{\rm{cov, abs}}(<R_{\rm{vir}})$ and use our zoom-in simulations to predict its values. 
For each simulation snapshot, we evaluate the covering fraction by extracting absorbers along one LOS for each grid pixel with impact parameter $<R_{\rm vir}$. 
To capture viewing angle variance, we compute covering fractions for three orthogonal sky projections for each snapshot. 
For each halo, we plot in the middle panel of Figure~\ref{three_panel} the average covering fraction and the estimated standard deviation among the projections using an error bar. 
For this figure we include all systems with $16.2 \leq \log{N_{\rm HI}} \leq 19$ to match the observational analysis of \cite{Lehner2013}. 
This panel shows that the LLS covering fraction versus halo mass at $z\leq1$ peaks broadly for $10^{10} M_\odot \lesssim \Mvir \lesssim 10^{12} M_\odot$. 
A similar (but quantitatively different) trend for LLS covering fractions peaking in relatively massive halos is seen in the FIRE simulations at $z=2-4$ \citep[][]{Faucher-Giguere2015, Faucher-Giguere2016}. 

In all snapshots, the variance in $f_{\rm{cov}}$ (the error bars in Figure~\ref{three_panel}) owing to different viewing angles is small compared to the variance arising from time variability and halo-to-halo scatter. 
To demonstrate the strong time variability of the covering fractions in individual halos, we plot in Figure \ref{f_cov_v_redshift} $f_{\rm cov}$ as a function of redshift for the six halos shown in Figure \ref{snapshot_summary}. 
The curves in Figure \ref{f_cov_v_redshift} are based on the 31 simulation snapshots between $z=1$ and $z=0$ analyzed in this paper for each simulation. 
The time-dependence of the covering fractions arises from the time-dependence of cosmological inflows inherited from the evolution of large-scale cosmic structure and from the bursty nature of star formation and galactic winds predicted by our simulations \citep[e.g.,][]{Muratov2015, Faucher-Giguere2015, Sparre2015}. 
In addition to the time variability in individual halos, there could be some differences in time-averaged covering fractions between different halos, for example connected to the environments of halos.
In \S \ref{sec:metallicity_distribution}, we perform a bootstrapping analysis to quantify the uncertainty in the metallicity distribution predicted using our set of zoom-in simulations arising from halo-to-halo scatter.

We fit our simulated covering fraction data to a function of halo mass with a redshift-dependent double power law of the form
\begin{equation}
f_{\rm cov}({\rm LLS}; <\Rvir) = \frac {2 f_{\rm cov, *} (1 + z)^\alpha}{(\Mvir/M_*)^{-\beta} + (\Mvir/M_{*})^{\gamma}}.
\end{equation}
Performing a least squares fit to all the $f_{\rm cov}$ data in the redshift interval $0 \le z \le 1$ (for $16.2 \leq \log \NHI \leq 19$)  results in the best-fit parameters $(f_{\rm cov, *},\log M_*,\beta,\gamma,\alpha)$ = $(0.012\pm0.001, 10.22\pm0.02, 3.9\pm0.1, 0.22\pm0.03, 2.5\pm0.2)$. 
The root mean square (r.m.s.) $\log{f_{\rm cov}}$ error for the fit is 0.34. 
The fit (plus and minimum the r.m.s. error) is plotted in the middle panel of Figure~\ref{three_panel} for $z=0$ and $z=1$. 
To match the LLS definition used in previous high-redshift FIRE papers \citep[][]{Faucher-Giguere2015, Faucher-Giguere2016}, we also fitted the covering fractions for $\log \NHI \ge 17.2$. 
The shape and redshift evolution of the best fit are consistent with the one above, so we held the $\log M_*$, $\beta$, $\gamma$, and $\alpha$ parameters fixed and solved for the best-fit normalization $f_{\rm cov, *}$ for the different $N_{\rm HI}$ cut. 
The best-fit normalization for $\log \NHI \ge 17.2$ is $f_{\rm cov, *}=0.0101\pm0.0003$, with an r.m.s. error of $\log{f_{\rm cov}}$ of 0.30. 
This is $20\%$ lower than for $16.2 \leq \log \NHI \leq 19$ because higher-column systems are generally rarer. 

The contribution of different halos to the cosmological incidence of absorbers can then be expressed in terms of functions describing the covering fraction as a function of redshift and halo mass, and the halo mass function:
\begin{equation}
w_{\rm{abs}}(z, \Mvir) = f_{\rm{cov, abs}}(<R_{\rm{vir}};~z,~M_{\rm h}) \frac{d^{2}N_{\rm{h}}}{dz~d\logMvir}.
\label{dN_LLS}
\end{equation}
For LLSs, $w_{\rm abs}$ is equivalent to $d^{2} N_{\rm LLS} / dz d\log{M_{\rm h}}$.  
We show $d^{2} N_{\rm LLS} / dz d\log{M_{\rm h}}$ predicted by our simulations at $z=0,~0.5,~$and 1 in the bottom panel of Figure \ref{three_panel}. 
The bottom panel shows that, at $z\leq1$, the majority of randomly-selected LLSs arise from halos in the mass range $10^{10} M_\odot \lesssim \Mvir \lesssim 10^{12} M_\odot$, where covering fractions peak. 

\begin{figure}
\begin{center}
\includegraphics[width=\linewidth, keepaspectratio]{{{totval_X_redshift_Y_f_cov_sparse_linear}}}
\caption{
Covering fractions for systems with $16.2 \le \log \NHI \le 19$ as a function of redshift for the six simulated halos shown in Figure~\ref{snapshot_summary}. 
The burstiness of star formation and galactic winds in our simulations drives the time variability of the covering fractions. 
}
\label{f_cov_v_redshift}
\end{center}
\end{figure}

\subsection{Cosmological incidence}
\label{l_z}

\begin{figure}
\begin{center}
\includegraphics[width=\linewidth, keepaspectratio]{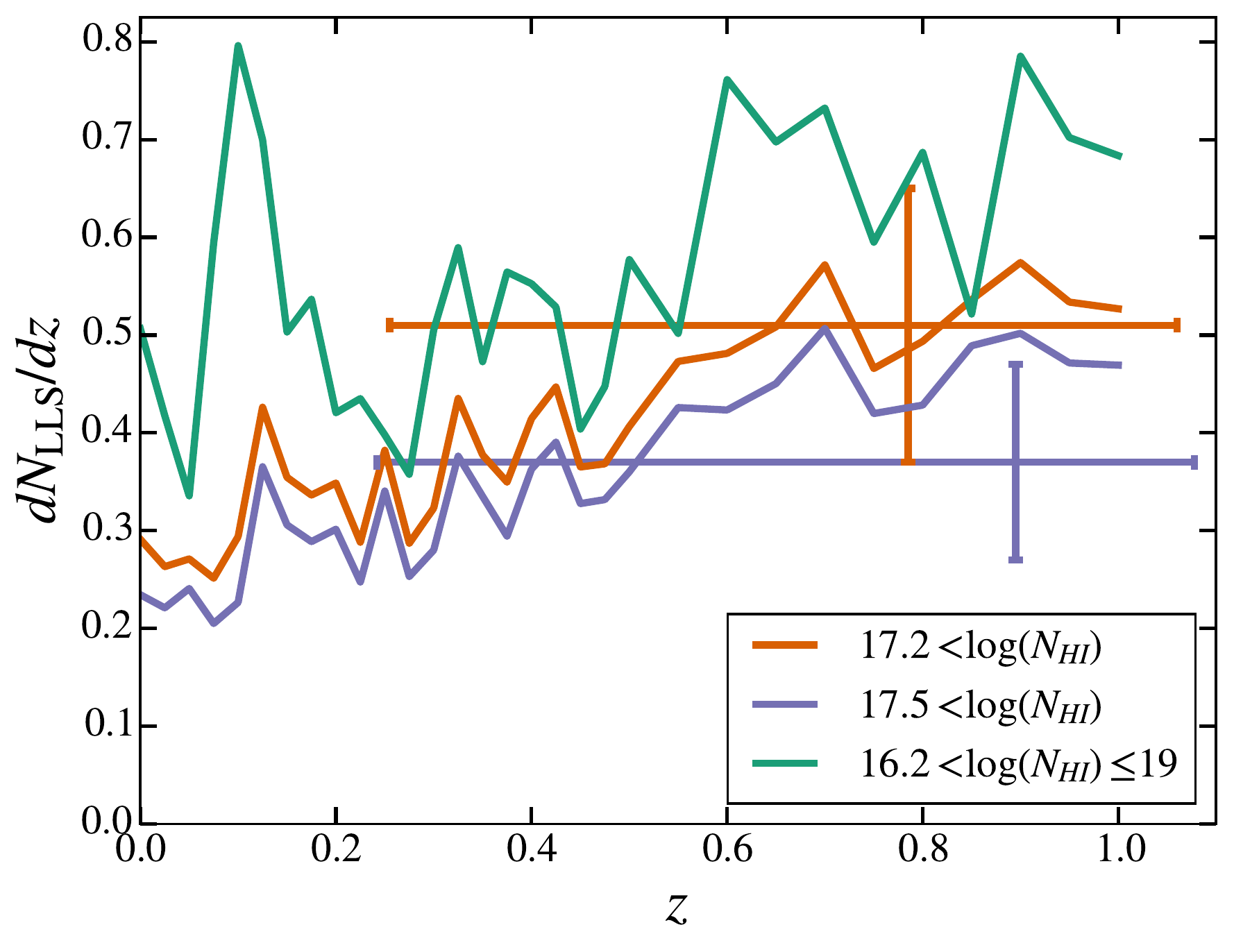}
\caption{Solid curves: The $z \leq 1$ cosmological incidence, $l(z)$, of LLSs predicted by our simulations for different HI column ranges ($10^{16.2} \le N_{\rm{H I}} \le 10^{19}$ in green, $N_{\rm{H I}} \ge 10^{17.2}$ in orange and $N_{\rm{H I}} \ge 10^{17.5}$ in blue). 
The error bars show observational measurements of $l(z)$ from the LLS census of \protect\cite{Ribaudo2011}. 
}
\label{l_z_plot}
\end{center}
\end{figure}

The total cosmological incidence of absorbers is obtained by integrating $w_{\rm abs}$ over halo mass:
\begin{equation}
l(z) \equiv \frac{dN_{\rm{abs}}}{dz} = \int d\logMvir  w_{\rm{abs}}(z, \Mvir). 
\end{equation}

We show the cosmological incidence of absorbers as a function of redshift for three different $N_{\rm HI}$ ranges in Figure \ref{l_z_plot}. 
The first range, $16.2 \leq \log{N_{\rm HI}} \leq 19$, is chosen to match the selection from \cite{Lehner2013}. 
The two other ranges, $\log{N_{\rm{H I}}} \ge 17.2$ and $\log{N_{\rm{H I}}} \ge 17.5$, match the observational measurements reported in \cite{Ribaudo2011}. 
The comparison with the observational compilation of \cite{Ribaudo2011} shows that convolving the covering fractions predicted by our zoom-in simulations with the halo mass function yields cosmological incidences of LLSs that are consistent with observations at $z \leq 1$. 
Overall, the simulated $l(z)$ increases slightly from $z=0$ to $z=1$; the significant redshift fluctuations apparent in Figure \ref{l_z_plot} reflect the strong time variability of covering fractions in individual simulations in our zoom-in sample (Fig. \ref{f_cov_v_redshift}).

\subsection{Column density distribution}
\label{diff_HI_dist}

\begin{figure}
\begin{center}
\includegraphics[width=\linewidth, keepaspectratio]{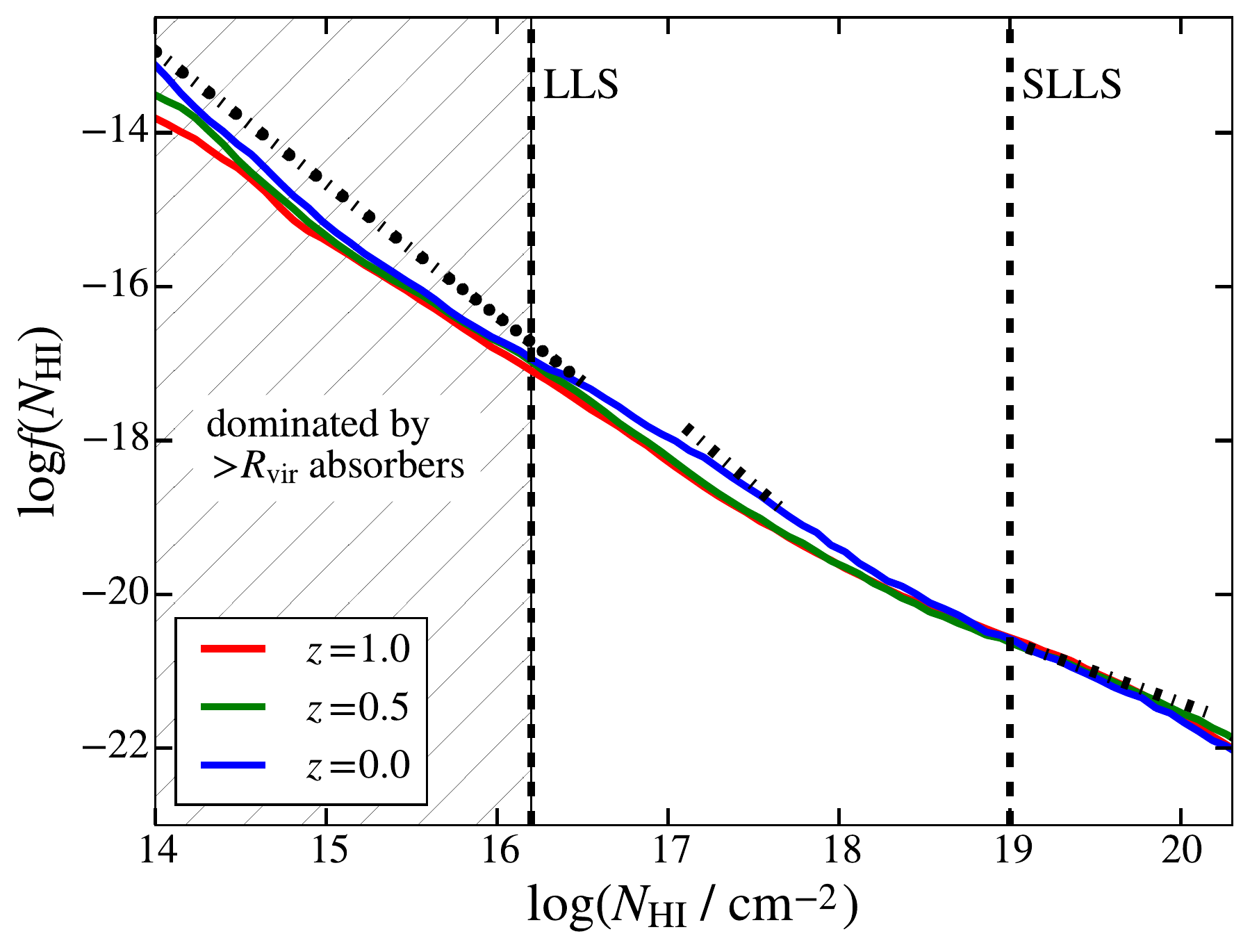}
\caption{H I column density distribution, $f(\NHI)$, predicted using our simulations at $z=0$, 0.5, and 1. The black dash-dotted lines show observational constraints from \citet{Ribaudo2011}. 
In the LLS and SLLS regimes, our simulations agree well with the observational constraints. 
For lower column systems, our simulations lie below the observational constraints by up to $\gtrsim0.5$ dex but this is because many such low-column absorbers arise outside the virial radius of galaxy halos and therefore are not accurately captured by our analysis, which focuses on $<R_{\rm vir}$ absorbers.
}
\label{diff_HI_dist_plot}
\end{center}
\end{figure}

Our zoom-in simulations can also be used to predict the column density distribution function of CGM absorbers, $f(N_{\rm{H I}})$, defined as the incidence of absorption systems per unit $N_{\rm{H I}}$ per unit absorption length, i.e.
\begin{equation}
f(N_{\rm{H I}}) \equiv \frac {d^2 N_{\rm{abs}}} {dN_{\rm{H I}} dX}.
\end{equation}
The absorption length, $X$, is defined such that the incidence of absorbers per absorption length, $l(X)$, is constant if the comoving surface area of the absorbers is constant across cosmic time \citep{Bahcall1969}:
\begin{equation}
l(X)dX = l(z)dz,
\end{equation}
where
\begin{equation}
dX = dz~(1 + z)^2 H_0 / H(z) .
\end{equation}
Using the procedure of the previous section to compute $l(z)$ for small column density intervals, we can compute $f(N_{\rm{H I}})$: 
\begin{equation}
f(N_{\rm{H I}}) \approx \frac{\Delta l(z)} {\Delta N_{\rm{H I}}}  \frac{dz}{dX},
\end{equation}
where $\Delta l(z)$ is the contribution to the incidence of absorbers for systems with column between $N_{\rm HI}$ and $N_{\rm HI}+\Delta N_{\rm HI}$ at redshift $z$. 

Figure~\ref{diff_HI_dist_plot} shows the column density distribution predicted by our simulations for $10^{14}~{\rm cm^{-2}}\leq N_{\rm HI} \leq 2\times10^{20}$ cm$^{-2}$. 
These predictions are compared to the observational constraints on the column density distribution from \cite{Ribaudo2011}. 
Overall, the predicted column density distribution agrees well with observations for LLSs and SLLSs but the simulations under-predict the column density distribution of lower-column systems by $\gtrsim0.5$ dex at $N_{\rm HI}\sim10^{15}$ cm$^{-2}$. 
Such a discrepancy at lower columns is expected since our analysis considers only absorbers that arise within the projected virial radius of main halos, i.e. it ignores absorbers that arise in the IGM outside dark matter halos.

\section{Relationships between HI column, metallicity and radial kinematics}
\label{hist_section}

We now quantify the relationships between the HI column density of absorbers, their metallicity, and their radial kinematics relative to central galaxies.  

For this analysis, we define ``cosmological histograms'' that enable us to study correlations between absorbers properties properly weighted by their cosmological incidence. For any prescribed $N_{\rm HI}$ interval, consider a random absorber (along the sight line to a random background quasar) with properties (e.g., HI column, metallicity, and radial velocity) described by the vector $\mathbf{Y}$. 
Following the formalism of \S \ref{abs_HI_stats}, the probability density function $\mathrm{P}(\mathbf{Y})$ is
\begin{equation}
\textrm{P}(\mathbf{Y}) =  \frac { \int \int  dz d\logMvir  \textrm{P}(\mathbf{Y} | z, \logMvir) w_{\rm{abs}}}  { \int \int dz d\logMvir  w_{\rm{abs}}}.
\label{mock_obs_hist}
\end{equation}
Here $\mathrm{P}(\mathbf{Y} | z, \logMvir)$ is the conditional probability density for $\mathbf{Y}$ given a redshift $z$ and halo mass $\Mvir$. 
In our case, $\mathrm{P}(\mathbf{Y} | z, \logMvir)$ is the joint probability density function of HI column density, metallicity, and radial velocity. 
We use three orthogonal sky projections for each snapshot in evaluating $\mathrm{P}(\mathbf{Y} | z, \logMvir)$, although the distribution varies relatively little for different viewing angles. 
Since $\mathbf{Y}$ is in general multi-variate, we can consider different 2D projections to quantify correlations between absorber attributes.

As a complement to the 2D cosmological histograms, we compute corresponding 2D histograms that quantify the mean halo mass contributing to different bins in parameter space:
\begin{align}
\langle \log{\matr{M}} \rangle = \frac { \int \int dz d\logMvir  \logMvir \mathrm{P}(\mathbf{Y} | z, \logMvir) w_{\rm abs}} 
{ \int \int dz d\logMvir \mathrm{P}(\mathbf{Y} | z, \logMvir) w_{\rm abs}}.
\label{avg_mass_diagram}
\end{align}
For each bin of $\mathbf{Y}$, this histogram gives the mean $\logMvir$ of halos contributing to the bin weighted by the cosmological incidence of absorbers. 

For any given column density interval (such as LLSs), the weighting by $w_{\rm abs}$ in equation (\ref{mock_obs_hist}) is derived such that the resulting 2D histogram represents the bivariate distribution of physical properties for absorbers in that column density interval along random sight lines in the Universe. 
Below, we are also interested in the metallicity distribution as a function of HI column. 
For this, we replace the weighting factor $w_{\rm abs}$ by simply $d^{2} N_{\rm h} / dz d\log{M_{\rm h}}$ in equations (\ref{mock_obs_hist}) and (\ref{avg_mass_diagram}).
This is because in this case $\textrm{P}(\mathbf{Y})$ is not restricted to a prescribed $N_{\rm HI}$ interval, so we need not include the $f_{\rm cov,abs}$ term in $w_{\rm abs}$.

All histograms are evaluated using all $0\leq z \leq1$ simulated data. 
Quantities are interpolated linearly with respect to $z$ and $\log{M_{\rm h}}$ to fill gaps between snapshots. 
Given the time variability of the CGM in individual halos (e.g., Fig. \ref{f_cov_v_redshift}) and our modest-sized sample of simulated halos, averaging over this full redshift interval is necessary to mitigate stochastic fluctuations. 
Furthermore, LLS measurements at $z\leq1$ are presently limited to a few dozen systems so their distributions of physical parameters are obtained by averaging over a similar redshift interval \citep[e.g.,][]{Lehner2013, 2016ApJ...831...95W}.

\subsection{Metallicity vs. HI column}

\begin{figure*}
\centering
\begin{minipage}{0.495\textwidth}
\centering
\includegraphics[width=\textwidth, keepaspectratio]{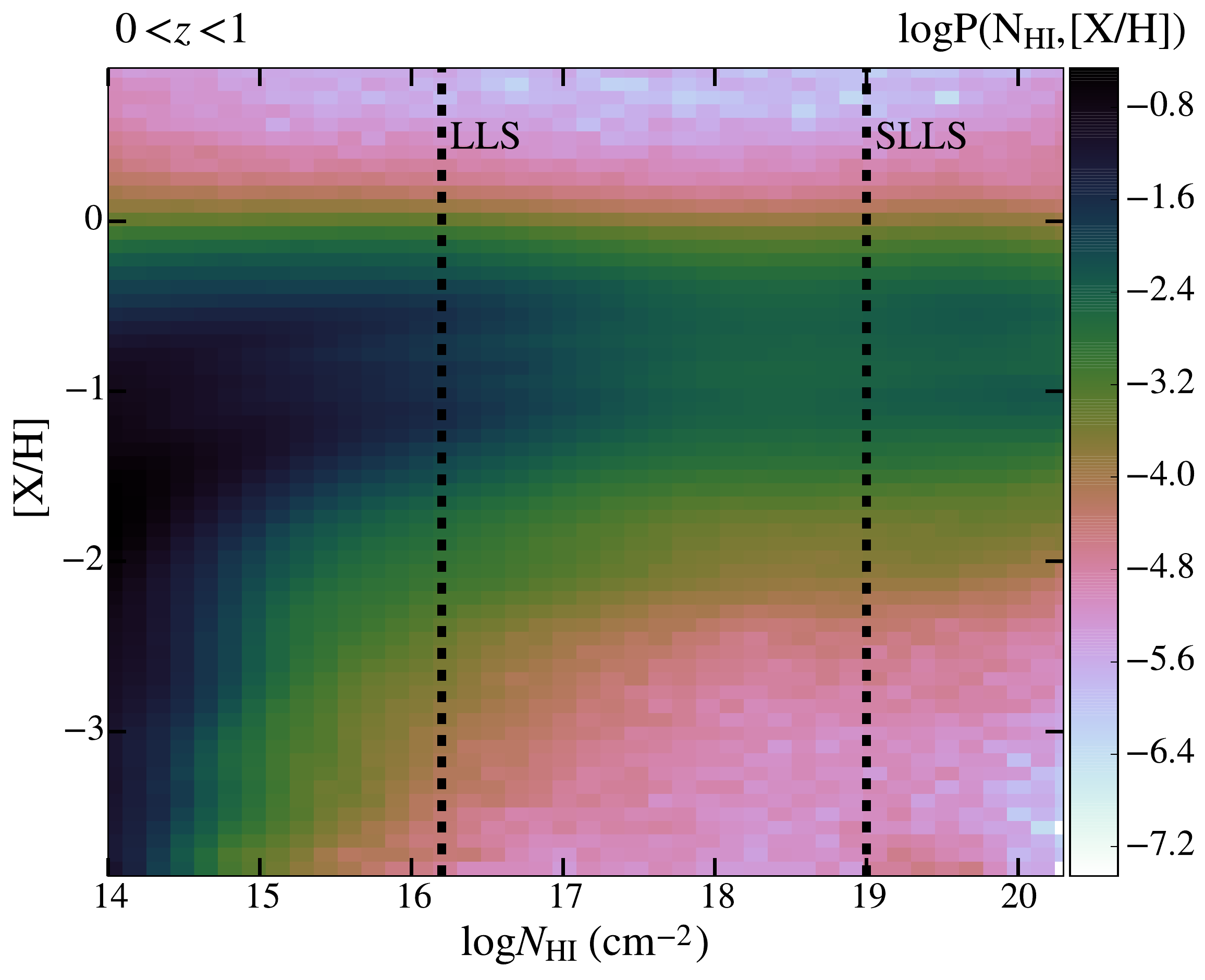}
\end{minipage} \hfill
\begin{minipage}{0.495\textwidth}
\centering
\includegraphics[width=\textwidth, keepaspectratio]{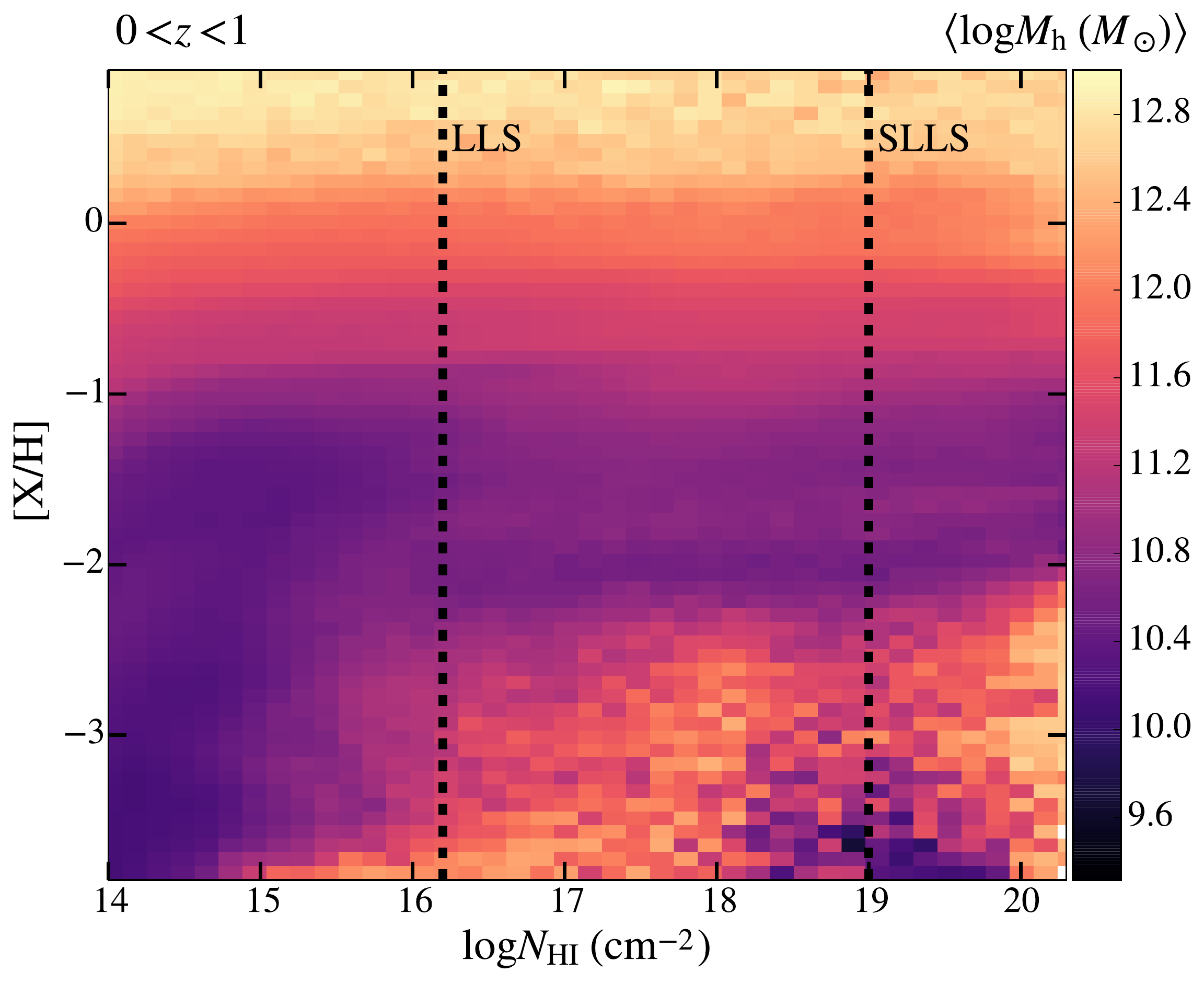}
\end{minipage}
\caption{\textbf{Left:} Cosmological 2D histogram for metallicity vs. HI column density from our simulations (representative of randomly selected LLSs). 
Simulated LLSs typically have a metallicity within 1 dex of the LLS mean metallicity $[X/H] \approx -0.9$. 
The metallicity distribution for lower column systems, $N_{\rm HI} \lesssim 10^{15}$ cm$^{-2}$, extends to lower metallicities. 
\textbf{Right:} 2D histogram showing the mean $\log{M_{\rm h}}$ contributing to different bins on the left hand side. 
Overall, the contribution of more massive halos is weighted toward higher metallicities. 
The color bars are logarithmically scaled. 
We include systems with $N_{\rm HI} < 10^{16.2}$ cm$^{-2}$ to provide information about their distribution integrated over galaxy halos, but caution that this distribution does not  include a contribution from absorbers arising outside the virial radius of halos.
}
\label{combined_hist2d_HI_Z}
\end{figure*}

Figure~\ref{combined_hist2d_HI_Z} shows the cosmological histogram for metallicity vs. HI column and the corresponding mean halo mass histogram. 
As the panel on the left reveals, simulated LLSs typically have a metallicity within 1 dex of the median metallicity ($[X/H]\approx-0.9$) of all LLSs. 
Lower column systems, $N_{\rm HI} \lesssim 10^{15}$ cm$^{-2}$ on the other hand, have a metallicity distribution that extends well below $[X/H]=-2$, consistent with many of these absorbers representing a tail of IGM gas that has not yet been significantly enriched by galaxies. 
In the next section, we show that most $[X/H]<-2$ absorbers have infall velocities and thus predominantly trace accretion of metal-poor intergalactic gas. 

Overall, the mean mass histogram on the right hand side of Figure \ref{combined_hist2d_HI_Z} reveals that more massive halos have a contribution weighted toward higher absorber metallicities. 
This overall trend is a CGM analog of the well-known mass-metallicity relation for galaxies \citep[e.g.,][]{Tremonti2004, Zahid2011, Peeples2014} and which the FIRE simulations broadly reproduce \citep[][]{Ma2016}. 
The $[X/H]<-2$ tail of the metallicity distribution is weighted toward $M_{\rm h} \lesssim 10^{11}$ M$_{\odot}$ halos. 

\subsection{Metallicity vs. radial kinematics}
\label{Z_v_vr}
\begin{figure*}
\centering
\begin{minipage}{0.495\textwidth}
\centering
\includegraphics[width=\textwidth, keepaspectratio]{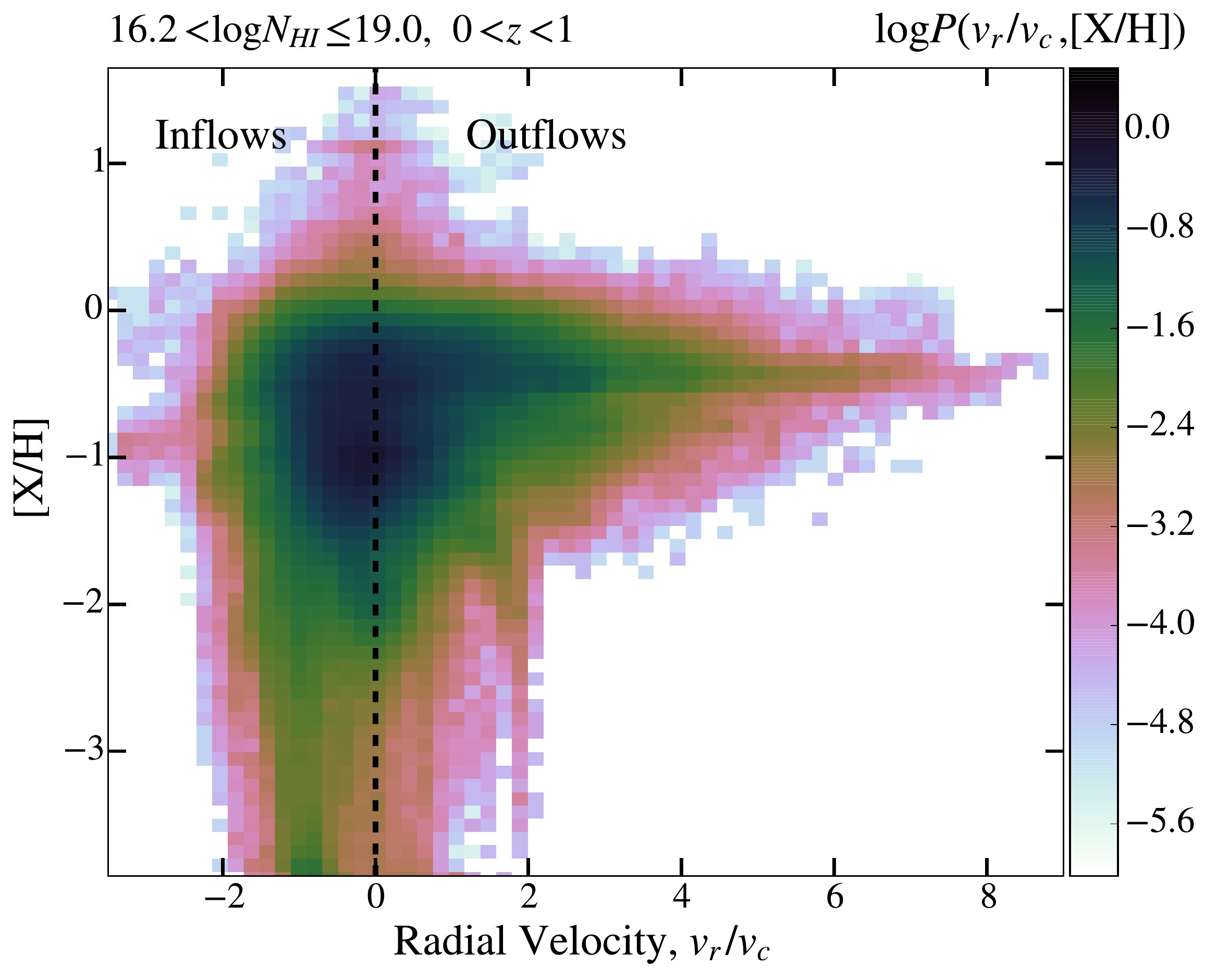}
\end{minipage} \hfill
\begin{minipage}{0.495\textwidth}
\centering
\includegraphics[width=\textwidth, keepaspectratio]{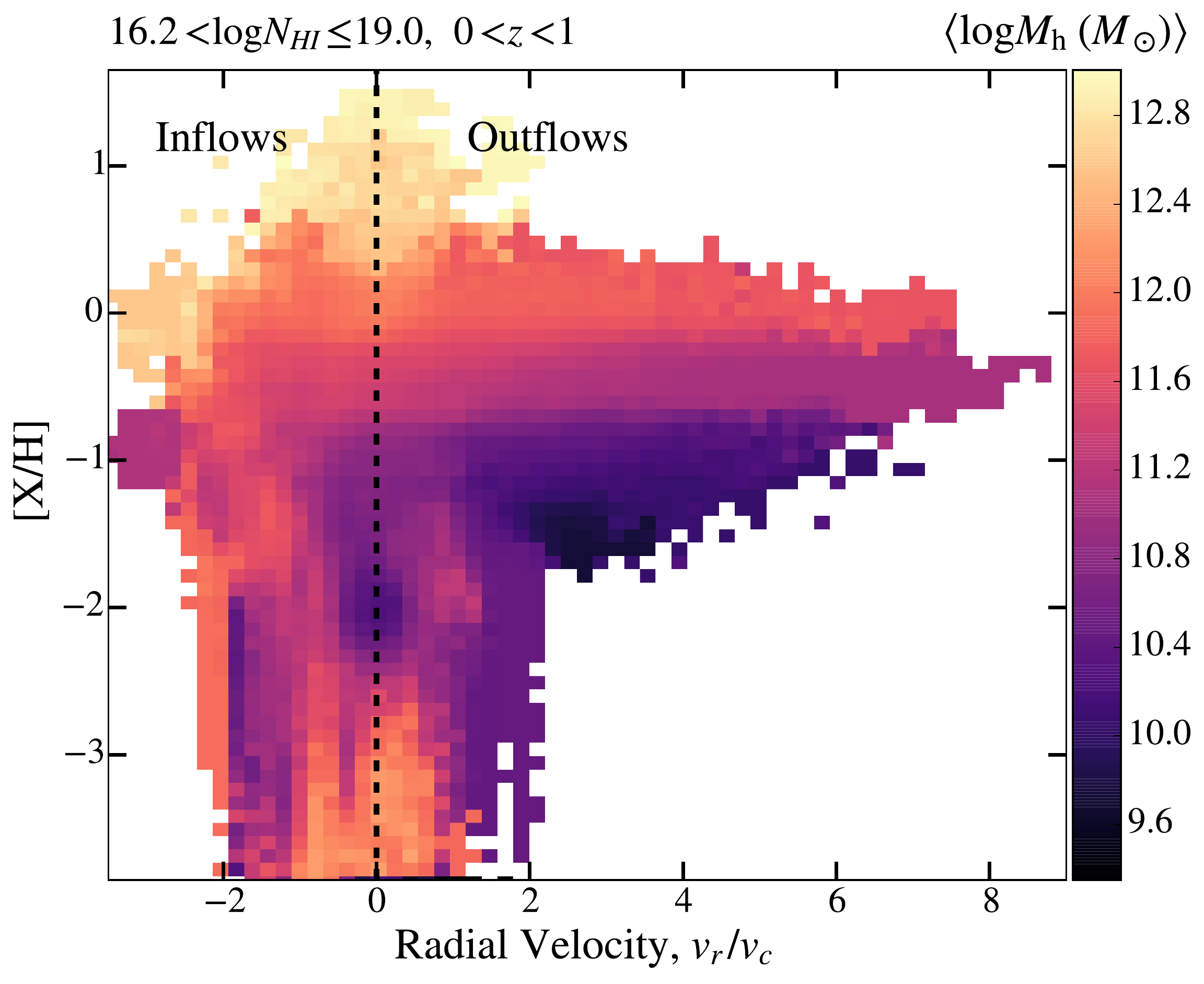}
\end{minipage}
\caption{\textbf{Left:} Cosmological 2D histogram for LLS metallicity vs. radial velocity relative to the central galaxy (representative of randomly selected LLSs). 
High velocity ($v_{\rm r}/v_{\rm c}>2$) outflows tend to have high metallicities $[X/H]\sim -0.5$ and very low metallicity ($[X/H]<-2$) LLSs tend to have infall radial velocities.
However, most LLSs occupy an intermediate central region in metallicity-radial velocity space. 
For these more typical LLSs, there is no clear trend between metallicity and radial kinematics. 
Metal-enriched inflows arise in the FIRE simulations as a result of galactic winds that efficiently recycle at low redshift. 
\textbf{Right:} Mean mass histogram showing the mean $\log{M_{\rm h}}$ contributing to different bins on the left hand side.
The color bars are logarithmically scaled. 
}
\label{combined_hist2d_vr_Z}
\end{figure*}

A primary motivation for our analysis is to understand how LLS metallicity can be used as a diagnostic of inflows and outflows. 
Figure~\ref{combined_hist2d_vr_Z} illustrates the relationship between absorber metallicity and its radial velocity relative to the central galaxy, $v_{\rm r}$, along with the corresponding mean mass histogram, for absorbers with $16.2 \leq \log{N_{\rm HI}} \leq 19$ (i.e., including partial LLSs). 
Positive radial velocities correspond to gas that is outflowing in an instantaneous sense, while negative radial velocities correspond to gas that is inflowing. 
In order to compare radial velocities for absorbers arising in a wide range of halo mass, velocities are normalized by halo circular velocity, $v_{\rm c} = \sqrt{G \Mvir/R_\textrm{vir}}$, which corresponds to $\sim(30,~130,~270)~ \textrm{km}/\textrm{s}$ for $\Mvir\sim (10^{10},~10^{12},~10^{13})$~\Msun. 

Figure \ref{combined_hist2d_vr_Z} shows that for cosmologically-selected LLSs in the redshift interval $0 \leq z \leq 1$ in our simulations:
\begin{enumerate}
\item Very high velocity ($v_{\rm r}/v_{\rm c}>2$) outflows tend to have high-metallicity $[X/H]\approx-0.5$. 
\item Very low metallicity ($[X/H]<-2$) LLSs tend to have radial velocities corresponding to inflows ($v_{\rm r}<0$).
\item Most LLSs have modest absolute velocity $|v_{\rm r}/v_{\rm c}|\leq2$. Overall, these LLSs have a broad metallicity distribution ranging from $[X/H] \approx -2$ to $[X/H] \approx 0$, with no clear trend with radial kinematics.
\end{enumerate}
Thus, our simulations suggest that very low metallicity LLSs with $[X/H]\lesssim-2$ generally trace infalling IGM gas \citep[consistent with observational interpretations of such low-metallicity absorbers, e.g.][]{Ribaudo2011a, Fumagalli2011b} but that metallicity alone cannot robustly distinguish between inflows and outflows for the majority of LLSs with higher metallicity.

The concentration of LLSs with $|v_{\rm r}/v_{\rm c}|\leq2$ is easy to understand since $v_{\rm c}$ is both the characteristic velocity of gas that is accelerated as it falls into halos from the IGM and the characteristic velocity of galactics winds in the FIRE simulations \citep[][]{Muratov2015}. 
The difficulty of associating metal-enriched LLSs with inflows vs. outflows is due in large part to the importance of wind recycling in our simulations \citep[for a previous analysis of how wind recycling shapes the galaxy stellar mass function, see][]{Oppenheimer2010}. 
When galactic winds recycle efficiently, metal-rich wind ejecta that initially have positive radial velocity later fall back onto the source galaxy with $v_{\rm r}<0$. 
\cite{Muratov2015} showed that the powerful galactic winds driven by high-redshift galaxies transform into galactic fountains by $z\sim0$ (or earlier) in massive galaxies in the FIRE simulations. 
\cite{Ma2016} furthermore demonstrated that the FIRE galaxies, except dwarfs with stellar mass $M_{\star}\lesssim 10^{7}$ M$_{\odot}$, retain most of the metals they have produced in their halos. 
\cite{AnglesAlcazar2016} confirm the importance of wind recycling in the FIRE simulations using a particle tracking analysis that traces the full history of baryons. 

As a check of our above result based on instantaneous radial kinematics that absorbers with $[X/H] \lesssim -2$ can generally be associated with the accretion of fresh gas from the IGM, we have used the particle tracking pipeline from \cite{AnglesAlcazar2016} and found the following:
in the main halo of our m12i simulation at $z=0.5$ nearly all the gas associated with LLSs and classified as fresh accretion based on its full time history has metallicity $[X/H] \lesssim -1$, with more than half having metallicity $[X/H] \lesssim -2$. 
On the other hand, nearly all the wind LLSs in this halo have metallicity $[X/H] \gtrsim -2$. 
We plan to more systematically extend this particle tracking analysis in future work.
We note that previous authors found using particle tracking that wind recycling can constitute a significant (and possibly dominant) fraction of the instantaneously infalling CGM gas at late times in other simulations as well \citep[][]{2014MNRAS.444.1260F, 2016ApJ...824...57C}. 
However, the magnitude of this effect depends on the properties of the (uncertain) wind models so previous results do not necessarily quantitatively apply to the FIRE simulations.

The ambiguities in using LLS metallicity to distinguish between inflows and outflows in the FIRE simulations at low redshift are in contrast to some earlier simulation analyses at high redshift ($z\sim3$), in which maps of the CGM appear to show a much stronger correlation between instantaneous radial kinematics and metallicity than the $z=0.5$ halos shown in Figure \ref{snapshot_summary} or than quantified in this section \citep[e.g.,][]{2012ApJ...760...50S}. 
Quantitatively, the analysis of a single main halo in the ErisMC simulation by \cite{2012ApJ...760...50S} revealed a factor of $\sim10$ difference in mean metallicity between instantaneously inflowing and outflowing halo gas at $z \geq 3$.
The difference between our results and the high redshift results of \cite{2012ApJ...760...50S} is likely at least partially attributable to the fact that both cosmological inflows and galactic winds are much stronger at high redshift. 
Furthermore, high redshift accretion of cool intergalactic gas dense enough to give rise to LLS absorption tends to be collimated in streams \citep[e.g.,][]{Keres2005, Dekel2009, Fumagalli2011, Faucher-Giguere2011} between which outflowing gas can propagate relatively unimpeded. 
By $z\sim0.5$, Figure \ref{snapshot_summary} shows that well-defined cool streams of infalling gas have largely disappeared in the simulations. 
We therefore caution that our present conclusions may not apply at $z>1$.

\begin{figure}
\begin{center}
\includegraphics[width=\linewidth, keepaspectratio]{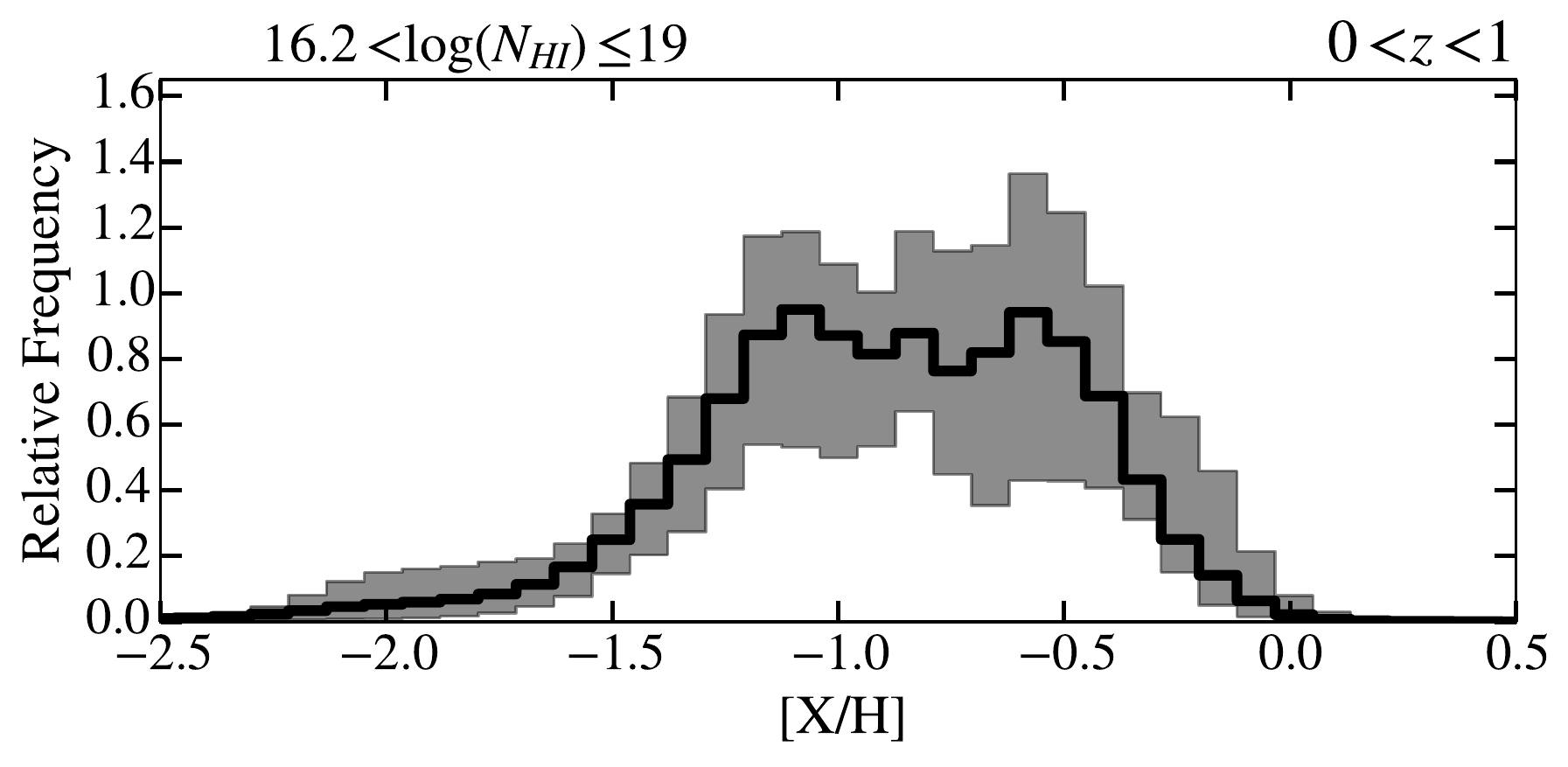}
\includegraphics[width=\linewidth, keepaspectratio]{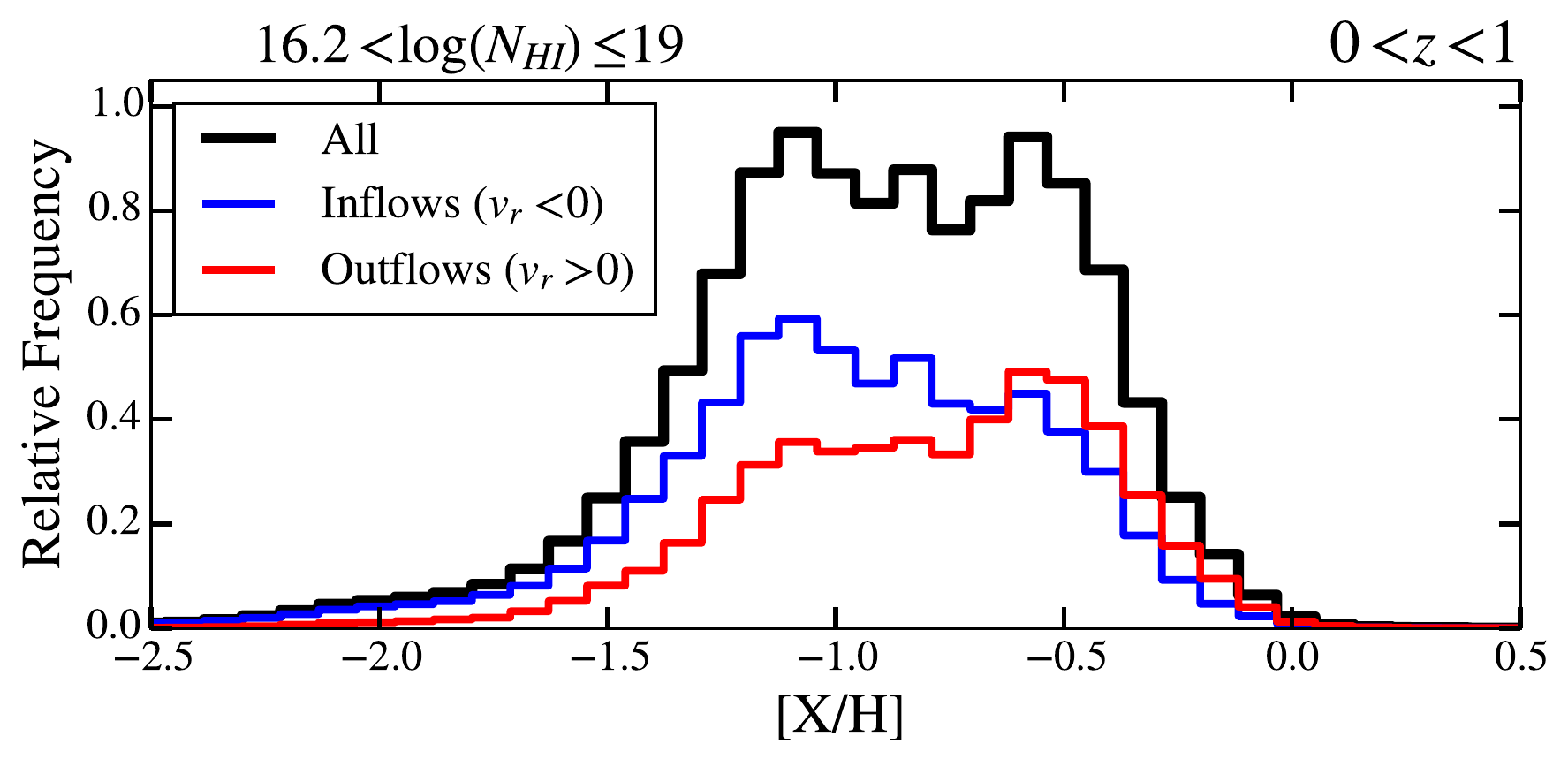}
\caption{\textbf{Top:} Overall metallicity distribution for $z \leq 1$ LLSs, averaging over halos of all masses (solid curve).
The shaded region shows the 95\% confidence interval estimated by sampling from the full simulation set with replacement. 
The multiple peaks apparent in the fiducial distribution are not significant. 
\textbf{Bottom:} Same LLS metallicity distribution as in the top panel (black) and divided into inflow (blue) and outflow (red) components based on instantaneous radial velocity. 
Outflowing LLSs have on average slightly higher metallicity than inflowing ones, but the two distributions overlap strongly, with no clean separation in metallicity.
}
\label{combined_hist_Z}
\end{center}
\end{figure}

\subsection{The metallicity distribution of Lyman limit systems}
\label{sec:metallicity_distribution}

To directly compare with the LLS metallicity distribution recently measured by \cite{Lehner2013}, we project the 2D histogram in Figure \ref{combined_hist2d_vr_Z} onto the metallicity axis. 
The result is shown in Figure \ref{combined_hist_Z}. 
The simulated metallicity distribution has a mean (standard deviation) $\langle{\rm [X/H]}\rangle =-0.9$ (0.4).

To test the statistical significance of features in the metallicity distribution, we performed a bootstrapping analysis. 
We recalculated the metallicity histogram 200 times, each time using a new sample of simulations drawn with replacement from our original set of simulations. 
The shaded region in the top panel of Figure \ref{combined_hist_Z} indicates the 95\% confidence interval for each metallicity bin. 
From this analysis, we conclude that apparent multiple peaks in the metallicity distribution computed using our entire simulation sample are consistent with statistical noise inherent to the finite number of simulations in our sample.
Thus, we do not find significant evidence for a multimodal metallicity distribution in our simulations. 
It is possible, though, that the simulated distribution has weak multimodal features within the noise of our analysis.

The lack of a significant metallicity bimodality in our simulations is in contrast with the possible metallicity bimodality observed by \cite{Lehner2013} and in the expanded sample of \cite{2016ApJ...831...95W}, which suggests a well-defined dip at $[X/H]\approx -1$. 
In our simulations, certain individual snapshots show rather well defined bimodal metallicity distributions (for example, the m11.9a snapshot shown in Fig. \ref{snapshot_summary}) with the low- and high-metallicity peaks roughly associated with inflows and galactic ejecta. 
However, inflows and outflows are highly time variable in our simulations and such features are transients. 
The main result of Figure \ref{combined_hist_Z} is that bimodal features in the metallicity distribution do not survive cosmological averaging over the broad range of halos in our simulations. 
This result is consistent with the broad range of halo mass contributing to the cosmological incidence of LLSs. 
Over the halo mass range $M_{\rm h} \sim 10^{10}-10^{12}$ M$_{\odot}$ that contributes most to the cosmological LLS incidence, the interstellar gas phase and stellar metallicities span a total range of $\sim2$ dex in our simulations (taking into account both the systematic increase with stellar mass and scatter around the mean mass-metallicity relation). 
These simulations are in good agreement with observational measurements of the mass-metallicity relations \citep[][]{Ma2016}.
Since the metallicity of galactic winds is similar to the metallicity of the interstellar medium \citep{2016arXiv160609252M}, the circum-galactic LLS metallicity distribution predicted by our simulations should therefore have a total width comparable to this. 
As a result, any clean bimodal ``dip'' in the metallicity distribution at a given halo mass (if present) may be expected to be substantially washed out in the cosmological average. 

Accretion of fresh gas from the IGM could in principle give rise to a distinct very low metallicity peak in the LLS metallicity distribution, but such a distinct component is not present in our simulated cosmological distribution. 
This is likely because even gas that is first accreting onto a central galaxy from the IGM tends to be pre-enriched by ejecta from surrounding lower-mass galaxies. 
Smooth IGM accretion generally originates from similar large-scale structures as infalling satellite galaxies, and we have previously noted the effects of galactic winds from satellites on the observational properties of infalling gas at high redshift \citep[][]{Faucher-Giguere2015, Faucher-Giguere2016}. 
  
In the updated $z \leq 1$ metallicity analysis of \cite{2016ApJ...831...95W}, the statistical evidence for a bimodality is strongest for the subsample of partial LLSs, with $16.2 \leq \log{N_{\rm HI}} \leq 17.2$. 
However, 15 out of 44 absorbers in that sample only have lower or upper limits on their metallicity. 
Allowing the lower limit absorbers to have arbitrarily high metallicity and the upper limit absorbers to have arbitrarily low metallicity could significantly flatten the distribution and potentially reduce the statistical significance of the dip at $[X/H] \approx -1$. 
It will be important to measure more precisely the metallicities of absorbers with only limits currently available (or to rigorously account for the limits in the statistical analysis) to more conclusively determine statistical significance of the observed dip. 
At $z>2$, the observational analyses of \cite{Fumagalli2015} and \cite{2016arXiv160802588L} reveal a broad unimodal LLS metallicity distribution. 

Another discrepancy between our simulations and the metallicity distributions measured by Lehner et al (2013) and \cite{2016ApJ...831...95W}, which cannot be explained by the presence of limits in the observational sample, is that our simulations contain significantly fewer very low metallicity LLSs than the observations. 
For example, our simulated metallicity distribution contains very few ($\approx 10\%$) LLSs with $[X/H]<-1.5$ but about half of \citeauthor{2016ApJ...831...95W}'s low-metallicity branch have such low metallicity. 
\citeauthor{2016ApJ...831...95W} explore systematic uncertainties in metallicity estimates from the assumed UV background model and find that using a UV background model with a harder spectrum due to a larger contribution from quasars would increase their inferred metallicities by up to $\sim0.6$ dex. This effect could explain part of the discrepancy at low metallicity if pushed to its extreme, but would introduce some tension at the high-metallicity end.

Differences in metal mixing may also contribute to the low-metallicity discrepancy. In our SPH simulations, metals are advected by particles but do not diffuse between particles, so that metal mixing occurring below the resolution limit is not captured. 
Underestimating subgrid metal mixing in the simulations should (if anything) overestimate the number of very low metallicity LLSs, rather than underestimate it. 
We have verified that the ``missing'' low-metallicity systems cannot be explained by the absence of subgrid metal mixing in our simulations by analyzing a re-simulation of our m12i halo that includes a subgrid turbulent metal diffusion model similar to \cite{2010MNRAS.407.1581S}. 
This re-simulation did not produce more $[X/H]<-1.5$ absorbers. 
On the other hand, if metals in the real CGM are poorly mixed, then cosmological simulations (both SPH and grid-based) could overestimate their mixing by forcing new metals to be distributed to at least one resolution element. 
In our simulations, metals produced by stellar evolution are distributed to the $\sim60$ gas particles in the SPH smoothing kernel, corresponding to a gas mass $>10^{6}$ M$_{\odot}$ for the m11.4a and m11.9a simulations whose parameters are listed in Table \ref{simulation_table}. 
In an observational analysis of $z\sim2.3$ CIV absorbers, \cite{2007MNRAS.379.1169S} concluded that those intergalactic absorbers likely arise from a large population of compact metal clumps with masses $M_{\rm clump} \sim 10^{2}$ M$_{\odot}$, which we are unable to resolve in our simulations. 
If the metals in $z \leq 1$ LLSs are similarly poorly mixed, then our simulations could overestimate the metallicity of many LLS sight lines within which metals in reality would not have had enough time to mix. 
This effect could potentially ``hide'' a population of low-metallicity LLSs. 

To further investigate the relationship between metallicity and inflows vs. outflows, we divide the total metallicity distribution into instantaneously inflowing ($v_{\rm r}<0$) and outflowing ($v_{\rm r}>0$) components in the bottom panel of Figure \ref{combined_hist_Z}. The instantaneously inflowing LLSs have a mean (standard deviation) $\langle{\rm [X/H]}\rangle =-1.0$ (0.4), while the instantaneously outflowing LLSs have a mean (standard deviation) $\langle{\rm [X/H]}\rangle =-0.8$ (0.4). 
Consistent with our results from the previous section, infalling LLSs are on average of slightly lower metallicity than outflowing ones, but the distributions are not cleanly separated in metallicity. This suggests that metallicity alone cannot be used as a diagnostic to distinguish instantaneous inflows from instantaneous outflows for LLSs with [X/H] $\gtrsim -2.0$, i.e. the majority of LLSs. However, as discussed in \S\ref{Z_v_vr}, the lowest metallicity systems with [X/H] $\lesssim -2$ do tend to be inflows. 

\section{Conclusions}
\label{conclusion}
We have used cosmological hydrodynamic zoom-in simulations from the FIRE project to investigate the physical properties of circum-galactic absorption at $z\leq 1$, with particular emphasis on Lyman limit systems and their use as diagnostics of cosmological inflows and galactic winds. 
The FIRE simulations self-consistently generate galactic winds from energy and momentum injection on the scale of star-forming regions by stellar feedback. 
We analyzed 14 simulations covering the halo mass range $M_\textrm{h} \sim 10^9-10^{13}$~\Msun~at $z=0$, which we convolved with the dark matter halo mass function to obtain cosmological statistics representative of absorbers randomly selected along unbiased sight lines to background quasars.
Our main conclusions are as follows:
\begin{enumerate}
\item When convolved with the dark matter halo mass function, the FIRE simulations are consistent with the observed cosmological incidence and HI column density distribution of Lyman limit systems at $z \leq 1$. 
\item The majority of HI-selected LLSs are associated with relatively massive halos in the mass range $10^{10} \lesssim \Mvir \lesssim 10^{12}$  \Msun. Dwarf halos with $M_{\rm h} \lesssim 10^{9.5}$ M$_{\odot}$ have extremely small ($\lesssim 10^{-4}$) LLS covering fractions within a projected virial radius. 
\item The LLS covering fractions of individual halos are strongly time variable. The strong variability results from a combination of time-variable inflows (including accreting satellites) and bursty outflows. 
\item Simulated LLSs typically have a metallicity within 1 dex of the mean metallicity ${\rm [X/H] } \approx-0.9$. 
The metallicity distribution of lower column systems, $N_{\rm HI} \lesssim 10^{15}$ cm$^{-2}$, extends well below $[X/H]=-2$. This is consistent with many of these low-column absorbers representing a tail of IGM gas that has not yet been significantly enriched by galaxies. 
\item LLSs with large radial outflow velocity ($v_{\rm r}/v_{\rm c}>2$) tend to have high metallicities $[X/H] \sim -0.5$, while very low metallicity ($[X/H]<-2$) LLSs tend to have infall radial velocities.
\item Most LLSs have moderate radial velocities ($|v_{\rm r}/v_{\rm c}|\leq2$). For these more common LLSs, there is no strong trend between metallicity and instantaneous radial kinematics. 
\item When separating LLSs in groups with $v_{\rm r}<0$ (inflows) and $v_{\rm r}>0$ (outflows), the inflowing LLSs have a slightly lower mean metallicity ($\langle{\rm [X/H]}\rangle = -1.0 $) than the outflowing LLSs ($\langle{\rm [X/H]}\rangle =-0.8 $). However, both distributions have a standard deviation of [X/H] $\approx 0.4$, causing them to overlap strongly.
\item Overall, we find no significant evidence for a bi-modality of LLS metallicity in our simulations. This result is in tension with observations that suggest two metallicity branches clearly separated at $[X/H] \approx -1$. 
\item The simulated metallicity distribution also lacks a population of low-metallicity LLSs ($[X/H] \lesssim -1.5$) that is prominent in observations, with only $\approx 10\%$ of all simulated LLSs having that low metallicity. The existence of such low-metallicity LLSs may indicate that metals are poorly mixed in the observed CGM on scales below the resolution of our simulations. 
\end{enumerate}
Overall, our simulations indicate that very low metallicity LLSs  
are predominantly associated with cosmological inflows at $z\leq1$, but that higher metallicity LLSs trace gas with roughly equal probability of having instantaneous inflow or outflow kinematics. 
This result is largely a consequence of the prevalence of gas recycling in the FIRE simulations, which causes a large fraction of metal-rich galactic wind ejecta to later fall back onto galaxies \citep[][]{AnglesAlcazar2016}.  
Thus, metallicity is a powerful diagnostic of pristine intergalactic inflows but in general cannot robustly distinguish between recent outflows and inflows of recycled gas. 
It will be interesting to sharpen this result by using particle tracking to more accurately identify the physical nature and history of gas elements in our simulations. 
Such an analysis may reveal that metallicity is a more powerful diagnostic of gas that has been previously processed by galaxies vs. not than of instantaneous inflows vs. instantaneous outflows.  
It will also be important to firm up the statistical significance of our analysis using a larger sample of simulated halos.

We also plan to extend our analysis of $z\leq1$ LLSs to the high-redshift Universe, where observational measurements of LLSs and their metallicities are also available \citep[][]{Lehner2014, Fumagalli2015, 2016arXiv160802588L}. 
At high redshift, many LLSs arise outside the virial radius of galaxy halos \citep[e.g.][]{Cooper2015} and the present zoom-in approach will not be adequate to study cosmological statistics. 
We plan to pursue full-volume cosmological simulations with the FIRE resolution and physics to address such IGM questions.
Another promising observational diagnostic of inflows and outflows that will be worthwhile to investigate using simulations is the distribution in azimuthal angle relative to the galactic disk plane of strong MgII absorbers \citep[e.g.,][]{Bordoloi2011, Bouche2012, Kacprzak2012, Lan2014}.

\section*{Acknowledgments}
The authors are grateful to Nicolas Lehner, Chris Wotta, Chris Howk, J. X. Prochaska, Cameron Liang, Joop Schaye, and Andrey Kravtsov for discussions regarding the observed Lyman limit system metallicity bimodality. 
We are also grateful to Benedikt Diemer for providing a set of Python cosmology modules used in this work.
ZH, CAFG, and DAA were supported by NSF through grants AST-1412836, AST-1517491 and DGE-0948017, by NASA through grant NNX15AB22G, and by STScI through grants HST-AR-14293.001-A and HST-GO-14268.022-A.
DK and TKC were supported in part by  NSF grant AST-1412153 and Cottrell Scholar Award from the Research Corporation for Science Advancement. 
Support for PFH was provided by an Alfred P. Sloan Research Fellowship, NASA ATP grant NNX14AH35G, and NSF grants AST-1411920 and AST-1455342. 
EQ was supported by NASA ATP grant 12-ATP-120183, a Simons Investigator award from the Simons Foundation, and the David and Lucile Packard Foundation. 
The simulations analyzed in this paper were run on XSEDE computational resources (allocations TG-AST120025, TG-AST130039, and TG-AST140023), on the NASA Pleiades cluster (allocation SMD-14-5189), on the Northwestern Quest computer cluster, and on the Caltech Zwicky computer cluster.

\appendix

\section{Convergence Properties}
\label{convergence_properties}

\begin{figure}
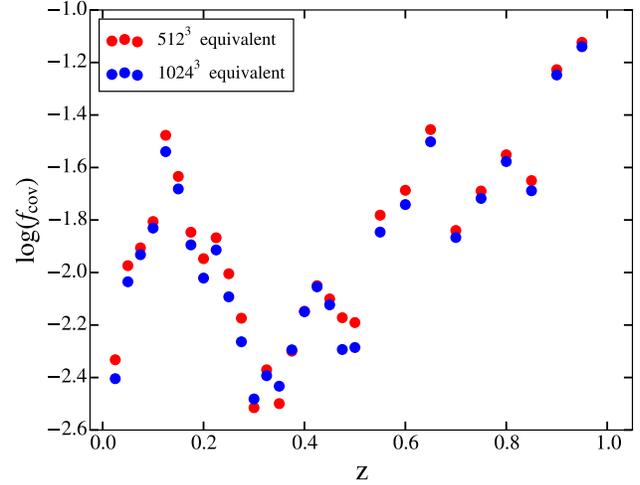

\begin{center}
\includegraphics[width=\linewidth, keepaspectratio]{{{f_cov_v_z_res_test}}}
\caption{
Grid resolution convergence for the LLS ($16.2 \leq \log \NHI \le 19$) covering fractions within a projected virial radius in the m12i simulation. We tested grids equivalent to our fiducial resolution and one higher resolution level ($512^3$ and $1024^3$ grid points on grids with side lengths of $2.4~\Rvir$), but spanning a single quadrant of the virial radius. The covering fractions are well converged for the fiducial $512^{3}$ grid resolution used in our analysis.
}
\label{HI_grid_convergence}
\end{center}
\end{figure}

\begin{figure}
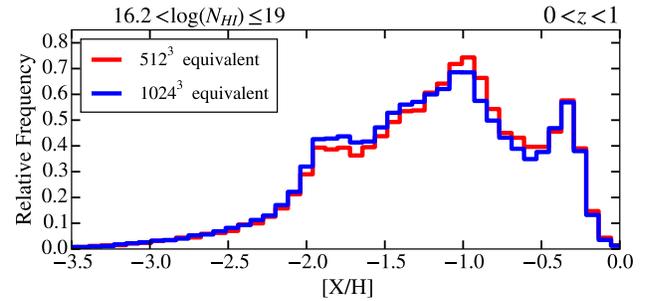

\begin{center}
\includegraphics[width=\linewidth, keepaspectratio]{{{Z_hist1d_grid_res}}}
\caption{
Same as Figure~\ref{HI_grid_convergence} but for the LLS metallicity distribution in the m12i simulation.
}
\label{Z_grid_convergence}
\end{center}
\end{figure}

To test our results for convergence with respect to grid resolution, we regridded a single quadrant of m12i at both fiducial resolution and at one higher resolution level over the full redshift range $0 \leq z \leq 1$. If we define the origin as the center of the m12i galaxy, our new cubic meshes have $256^3$ and $512^3$ grid points over a volume with $(x_{\rm min}, y_{\rm min}, z_{\rm min}) = (0, 0, 0)$ and $(x_{\rm max}, y_{\rm max}, z_{\rm max}) = (1.2 \Rvir, 1.2 \Rvir, 1.2 \Rvir)$. This produces grids with $512^{3}$ and $1,024^{3}$ equivalent resolution over the full halo. The results of our grid resolution convergence tests are shown in Figure~\ref{HI_grid_convergence} for LLS covering fractions and in Figure~\ref{Z_grid_convergence} for the metallicity distribution. Both the covering fractions and the metallicity distribution are well converged with respect to grid resolution. This is expected since the SPH smoothing lengths of LLS gas are comparable to the grid cell size at the fiducial grid resolution used for our analysis. 
We note that the LLS covering fractions plotted in Figure \ref{HI_grid_convergence} for the m12i simulation are somewhat more time variable than in Figure \ref{f_cov_v_redshift} in the main text because here we average over only a single quadrant of the simulation, whereas Figure \ref{f_cov_v_redshift} averages over the entire projected virial area.

\begin{figure}
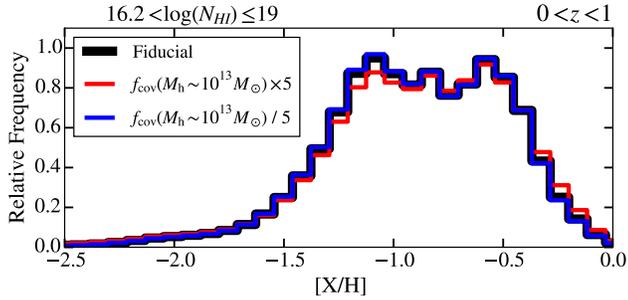

\begin{center}
\includegraphics[width=\linewidth, keepaspectratio]{{{Z_hist1d_mass_boosted}}}
\caption{
Overall cosmological LLS metallicity distribution recomputed by artificially increasing (red) and decreasing (blue) by a factor of 5 the weights of the most massive halos (m13 and MFz0\_A2), which are simulated with coarser mass resolution. 
The good agreement with the fiducial distribution (black) indicates that these contribute relatively little to the cosmological LLS metallicity distribution. The lower resolution of these simulations is thus unlikely to significantly affect our results.}
\label{massive_convergence}
\end{center}
\end{figure}

The most massive simulated halos in our sample, m13 and MFz0\_A2, were run at one resolution level lower (gas particles masses $m_{\rm b}\approx (3-4)\times10^{5}$ M$_{\odot}$) than the rest of our simulation sample. To test whether our results are sensitive to possible resolution effects in these simulations, we artificially varied their LLS covering fractions by a factor of 5 up and down and re-computed the cosmological metallicity distribution. Figure~\ref{massive_convergence} shows the results. Even if these massive halos were subject to significant resolution limitations, our main results would be affected only very slightly. This is because halos in this mass range are rare and contribute negligibly to global LLS statistics.

\bibliography{ref} 

\begin{thebibliography}{}

\bibitem[\protect\citeauthoryear{{Agertz}, {Moore}, {Stadel}, {Potter},
  {Miniati}, {Read}, {Mayer}, {Gawryszczak}, {Kravtsov}, {Nordlund}, {Pearce},
  {Quilis}, {Rudd}, {Springel}, {Stone}, {Tasker}, {Teyssier}, {Wadsley} \&
  {Walder}}{{Agertz} et~al.}{2007}]{Agertz2007}
{Agertz} O.,  {Moore} B.,  {Stadel} J.,  {Potter} D.,  {Miniati} F.,  {Read}
  J.,  {Mayer} L.,  {Gawryszczak} A.,  {Kravtsov} A.,  {Nordlund} {\AA}.,
  {Pearce} F.,  {Quilis} V.,  {Rudd} D.,  {Springel} V.,  {Stone} J.,  {Tasker}
  E.,  {Teyssier} R.,  {Wadsley} J.,    {Walder} R.,  2007, \mnras, 380, 963

\bibitem[\protect\citeauthoryear{{Aguirre}, {Hernquist}, {Schaye}, {Weinberg},
  {Katz} \& {Gardner}}{{Aguirre} et~al.}{2001}]{Aguirre2001}
{Aguirre} A.,  {Hernquist} L.,  {Schaye} J.,  {Weinberg} D.~H.,  {Katz} N.,
  {Gardner} J.,  2001, \apj, 560, 599

\bibitem[\protect\citeauthoryear{Altay, Theuns, Schaye, Booth \& Vecchia}{Altay
  et~al.}{2013}]{Altay2013}
Altay G.,  Theuns T.,  Schaye J.,  Booth C.~M.,    Vecchia C.~D.,  2013, MNRAS,
  436, 2689

\bibitem[\protect\citeauthoryear{{Altay}, {Theuns}, {Schaye}, {Crighton} \&
  {Dalla Vecchia}}{{Altay} et~al.}{2011}]{Altay2011}
{Altay} G.,  {Theuns} T.,  {Schaye} J.,  {Crighton} N.~H.~M.,    {Dalla
  Vecchia} C.,  2011, \apjl, 737, L37

\bibitem[\protect\citeauthoryear{{Angl{\'e}s-Alc{\'a}zar}, {Dav{\'e}},
  {{\"O}zel} \& {Oppenheimer}}{{Angl{\'e}s-Alc{\'a}zar}
  et~al.}{2014}]{AnglesAlcazar2014}
{Angl{\'e}s-Alc{\'a}zar} D.,  {Dav{\'e}} R.,  {{\"O}zel} F.,    {Oppenheimer}
  B.~D.,  2014, \apj, 782, 84

\bibitem[\protect\citeauthoryear{{Angl{\'e}s-Alc{\'a}zar},
  {Faucher-Gigu{\`e}re}, {Kere{\v s}}, {Hopkins}, {Quataert} \&
  {Murray}}{{Angl{\'e}s-Alc{\'a}zar} et~al.}{2016}]{AnglesAlcazar2016}
{Angl{\'e}s-Alc{\'a}zar} D.,  {Faucher-Gigu{\`e}re} C.-A.,  {Kere{\v s}} D.,
  {Hopkins} P.~F.,  {Quataert} E.,    {Murray} N.,  2016, ArXiv e-prints

\bibitem[\protect\citeauthoryear{Asplund, Grevesse, Sauval \& Scott}{Asplund
  et~al.}{2010}]{Asplund2010}
Asplund M.,  Grevesse N.,  Sauval A.~J.,    Scott P.,  2010, Astrophys. Space
  Sci., 328, 179

\bibitem[\protect\citeauthoryear{Bahcall \& Peebles}{Bahcall \&
  Peebles}{1969}]{Bahcall1969}
Bahcall J.~N.,  Peebles P. J.~E.,  1969, ApJ, 156, L7

\bibitem[\protect\citeauthoryear{{Barnes}}{{Barnes}}{2012}]{Barnes2012}
{Barnes} J.~E.,  2012, \mnras, 425, 1104

\bibitem[\protect\citeauthoryear{Bauermeister, Blitz \& Ma}{Bauermeister
  et~al.}{2010}]{Bauermeister2010}
Bauermeister A.,  Blitz L.,    Ma C.-P.,  2010, ApJ, 717, 323

\bibitem[\protect\citeauthoryear{{Bird}, {Vogelsberger}, {Sijacki},
  {Zaldarriaga}, {Springel} \& {Hernquist}}{{Bird} et~al.}{2013}]{Bird2013}
{Bird} S.,  {Vogelsberger} M.,  {Sijacki} D.,  {Zaldarriaga} M.,  {Springel}
  V.,    {Hernquist} L.,  2013, \mnras, 429, 3341

\bibitem[\protect\citeauthoryear{{Booth}, {Schaye}, {Delgado} \& {Dalla
  Vecchia}}{{Booth} et~al.}{2012}]{Booth2012}
{Booth} C.~M.,  {Schaye} J.,  {Delgado} J.~D.,    {Dalla Vecchia} C.,  2012,
  \mnras, 420, 1053

\bibitem[\protect\citeauthoryear{{Bordoloi et al.}}{{Bordoloi et
  al.}}{2011}]{Bordoloi2011}
{Bordoloi et al.} 2011, \apj, 743, 10

\bibitem[\protect\citeauthoryear{{Bouch{\'e}}, {Hohensee}, {Vargas},
  {Kacprzak}, {Martin}, {Cooke} \& {Churchill}}{{Bouch{\'e}}
  et~al.}{2012}]{Bouche2012}
{Bouch{\'e}} N.,  {Hohensee} W.,  {Vargas} R.,  {Kacprzak} G.~G.,  {Martin}
  C.~L.,  {Cooke} J.,    {Churchill} C.~W.,  2012, \mnras, 426, 801

\bibitem[\protect\citeauthoryear{{Brooks}, {Governato}, {Quinn}, {Brook} \&
  {Wadsley}}{{Brooks} et~al.}{2009}]{Brooks2009}
{Brooks} A.~M.,  {Governato} F.,  {Quinn} T.,  {Brook} C.~B.,    {Wadsley} J.,
  2009, \apj, 694, 396

\bibitem[\protect\citeauthoryear{{Bryan} \& {Norman}}{{Bryan} \&
  {Norman}}{1998}]{Bryan1998}
{Bryan} G.~L.,  {Norman} M.~L.,  1998, \apj, 495, 80

\bibitem[\protect\citeauthoryear{Chan, Kere{\v{s}}, O{\~{n}}orbe, Hopkins,
  Muratov, Faucher-Gigu{\`{e}}re \& Quataert}{Chan et~al.}{2015}]{Chan2015}
Chan T.~K.,  Kere{\v{s}} D.,  O{\~{n}}orbe J.,  Hopkins P.~F.,  Muratov a.~L.,
  Faucher-Gigu{\`{e}}re C.~a.,    Quataert E.,  2015, MNRAS, 454, 2981

\bibitem[\protect\citeauthoryear{{Christensen}, {Dav{\'e}}, {Governato},
  {Pontzen}, {Brooks}, {Munshi}, {Quinn} \& {Wadsley}}{{Christensen}
  et~al.}{2016}]{2016ApJ...824...57C}
{Christensen} C.~R.,  {Dav{\'e}} R.,  {Governato} F.,  {Pontzen} A.,  {Brooks}
  A.,  {Munshi} F.,  {Quinn} T.,    {Wadsley} J.,  2016, \apj, 824, 57

\bibitem[\protect\citeauthoryear{{Cicone et al.}}{{Cicone et
  al.}}{2014}]{Cicone2014}
{Cicone et al.} 2014, \aap, 562, A21

\bibitem[\protect\citeauthoryear{Cooper, Simcoe, Cooksey, O'Meara \&
  Torrey}{Cooper et~al.}{2015}]{Cooper2015}
Cooper T.~J.,  Simcoe R.~A.,  Cooksey K.~L.,  O'Meara J.~M.,    Torrey P.,
  2015, ApJ, 812, 58

\bibitem[\protect\citeauthoryear{{Dekel} \& {Birnboim}}{{Dekel} \&
  {Birnboim}}{2006}]{Dekel2006}
{Dekel} A.,  {Birnboim} Y.,  2006, \mnras, 368, 2

\bibitem[\protect\citeauthoryear{{Dekel}, {Birnboim}, {Engel}, {Freundlich},
  {Goerdt}, {Mumcuoglu}, {Neistein}, {Pichon}, {Teyssier} \& {Zinger}}{{Dekel}
  et~al.}{2009}]{Dekel2009}
{Dekel} A.,  {Birnboim} Y.,  {Engel} G.,  {Freundlich} J.,  {Goerdt} T.,
  {Mumcuoglu} M.,  {Neistein} E.,  {Pichon} C.,  {Teyssier} R.,    {Zinger} E.,
   2009, \nat, 457, 451

\bibitem[\protect\citeauthoryear{{Faucher-Gigu{\`e}re}, {Feldmann}, {Quataert},
  {Kere{\v s}}, {Hopkins} \& {Murray}}{{Faucher-Gigu{\`e}re}
  et~al.}{2016}]{Faucher-Giguere2016}
{Faucher-Gigu{\`e}re} C.-A.,  {Feldmann} R.,  {Quataert} E.,  {Kere{\v s}} D.,
  {Hopkins} P.~F.,    {Murray} N.,  2016, \mnras, 461, L32

\bibitem[\protect\citeauthoryear{Faucher-Gigu{\`{e}}re, Hopkins, Kere, Muratov,
  Quataert \& Murray}{Faucher-Gigu{\`{e}}re et~al.}{2015}]{Faucher-Giguere2015}
Faucher-Gigu{\`{e}}re C.-A.,  Hopkins P.~F.,  Kere D.,  Muratov A.~L.,
  Quataert E.,    Murray N.,  2015, MNRAS, 449, 987

\bibitem[\protect\citeauthoryear{Faucher-Gigu{\`{e}}re \&
  Kere{\v{s}}}{Faucher-Gigu{\`{e}}re \&
  Kere{\v{s}}}{2011}]{Faucher-Giguere2011}
Faucher-Gigu{\`{e}}re C.~A.,  Kere{\v{s}} D.,  2011, MNRAS, 412, 118

\bibitem[\protect\citeauthoryear{{Faucher-Gigu{\`e}re}, {Kere{\v s}} \&
  {Ma}}{{Faucher-Gigu{\`e}re} et~al.}{2011}]{FaucherGiguere2011}
{Faucher-Gigu{\`e}re} C.-A.,  {Kere{\v s}} D.,    {Ma} C.-P.,  2011, \mnras,
  417, 2982

\bibitem[\protect\citeauthoryear{Faucher-Gigu{\`{e}}re, Lidz, Zaldarriaga \&
  Hernquist}{Faucher-Gigu{\`{e}}re et~al.}{2009}]{Faucher-Giguere2009}
Faucher-Gigu{\`{e}}re C.~a.,  Lidz A.,  Zaldarriaga M.,    Hernquist L.,  2009,
  ApJ, 703, 1416

\bibitem[\protect\citeauthoryear{{Feldmann}, {Quataert}, {Hopkins},
  {Faucher-Gigu{\`e}re} \& {Kere{\v s}}}{{Feldmann}
  et~al.}{2016}]{2016arXiv161002411F}
{Feldmann} R.,  {Quataert} E.,  {Hopkins} P.~F.,  {Faucher-Gigu{\`e}re} C.-A.,
    {Kere{\v s}} D.,  2016, ArXiv e-prints

\bibitem[\protect\citeauthoryear{{Ford}, {Dav{\'e}}, {Oppenheimer}, {Katz},
  {Kollmeier}, {Thompson} \& {Weinberg}}{{Ford}
  et~al.}{2014}]{2014MNRAS.444.1260F}
{Ford} A.~B.,  {Dav{\'e}} R.,  {Oppenheimer} B.~D.,  {Katz} N.,  {Kollmeier}
  J.~A.,  {Thompson} R.,    {Weinberg} D.~H.,  2014, \mnras, 444, 1260

\bibitem[\protect\citeauthoryear{Fox, Lehner, Tumlinson, Howk, Tripp,
  Prochaska, O'Meara, Werk, Bordoloi, Katz, Oppenheimer \& Dav{\'{e}}}{Fox
  et~al.}{2013}]{Fox2013}
Fox A.~J.,  Lehner N.,  Tumlinson J.,  Howk J.~C.,  Tripp T.~M.,  Prochaska
  J.~X.,  O'Meara J.~M.,  Werk J.~K.,  Bordoloi R.,  Katz N.,  Oppenheimer
  B.~D.,    Dav{\'{e}} R.,  2013, ApJ, 778, 187

\bibitem[\protect\citeauthoryear{Fumagalli, Hennawi, Prochaska, Kasen, Dekel,
  Ceverino \& Primack}{Fumagalli et~al.}{2014}]{Fumagalli2014}
Fumagalli M.,  Hennawi J.~F.,  Prochaska J.~X.,  Kasen D.,  Dekel A.,  Ceverino
  D.,    Primack J.,  2014, ApJ, 780, 74

\bibitem[\protect\citeauthoryear{{Fumagalli}, {O'Meara} \&
  {Prochaska}}{{Fumagalli} et~al.}{2011}]{Fumagalli2011b}
{Fumagalli} M.,  {O'Meara} J.~M.,    {Prochaska} J.~X.,  2011, Science, 334,
  1245

\bibitem[\protect\citeauthoryear{{Fumagalli}, {O'Meara} \&
  {Prochaska}}{{Fumagalli} et~al.}{2016}]{Fumagalli2015}
{Fumagalli} M.,  {O'Meara} J.~M.,    {Prochaska} J.~X.,  2016, \mnras, 455,
  4100

\bibitem[\protect\citeauthoryear{{Fumagalli}, {O'Meara}, {Prochaska} \&
  {Worseck}}{{Fumagalli} et~al.}{2013}]{Fumagalli2013b}
{Fumagalli} M.,  {O'Meara} J.~M.,  {Prochaska} J.~X.,    {Worseck} G.,  2013,
  \apj, 775, 78

\bibitem[\protect\citeauthoryear{Fumagalli, Prochaska, Kasen, Dekel, Ceverino
  \& Primack}{Fumagalli et~al.}{2011}]{Fumagalli2011}
Fumagalli M.,  Prochaska J.~X.,  Kasen D.,  Dekel A.,  Ceverino D.,    Primack
  J.~R.,  2011, MNRAS, 418, 1796

\bibitem[\protect\citeauthoryear{{Goerdt}, {Dekel}, {Sternberg}, {Gnat} \&
  {Ceverino}}{{Goerdt} et~al.}{2012}]{Goerdt2012}
{Goerdt} T.,  {Dekel} A.,  {Sternberg} A.,  {Gnat} O.,    {Ceverino} D.,  2012,
  \mnras, 424, 2292

\bibitem[\protect\citeauthoryear{{Greene}, {Zakamska}, {Ho} \&
  {Barth}}{{Greene} et~al.}{2011}]{Greene2011}
{Greene} J.~E.,  {Zakamska} N.~L.,  {Ho} L.~C.,    {Barth} A.~J.,  2011, \apj,
  732, 9

\bibitem[\protect\citeauthoryear{{Gurvich}, {Burkhart} \& {Bird}}{{Gurvich}
  et~al.}{2016}]{2016arXiv160803293G}
{Gurvich} A.,  {Burkhart} B.,    {Bird} S.,  2016, arXiv:1608.03293

\bibitem[\protect\citeauthoryear{{Heckman}, {Lehnert}, {Strickland} \&
  {Armus}}{{Heckman} et~al.}{2000}]{Heckman2000}
{Heckman} T.~M.,  {Lehnert} M.~D.,  {Strickland} D.~K.,    {Armus} L.,  2000,
  \apjs, 129, 493

\bibitem[\protect\citeauthoryear{{Hernquist}, {Katz}, {Weinberg} \&
  {Miralda-Escud{\'e}}}{{Hernquist} et~al.}{1996}]{Hernquist1996}
{Hernquist} L.,  {Katz} N.,  {Weinberg} D.~H.,    {Miralda-Escud{\'e}} J.,
  1996, \apjl, 457, L51

\bibitem[\protect\citeauthoryear{{Hopkins}}{{Hopkins}}{2013}]{Hopkins2013d}
{Hopkins} P.~F.,  2013, \mnras, 428, 2840

\bibitem[\protect\citeauthoryear{Hopkins}{Hopkins}{2015}]{Hopkins2015}
Hopkins P.~F.,  2015, MNRAS, 450, 53

\bibitem[\protect\citeauthoryear{Hopkins, Keres, Onorbe, Faucher-Giguere,
  Quataert, Murray \& Bullock}{Hopkins et~al.}{2014}]{Hopkins2014}
Hopkins P.~F.,  Keres D.,  Onorbe J.,  Faucher-Giguere C.-A.,  Quataert E.,
  Murray N.,    Bullock J.~S.,  2014, MNRAS, 445, 581

\bibitem[\protect\citeauthoryear{Jones, Stark \& Ellis}{Jones
  et~al.}{2012}]{Jones2012}
Jones T.,  Stark D.~P.,    Ellis R.~S.,  2012, ApJ, 751, 51

\bibitem[\protect\citeauthoryear{{Kacprzak}, {Churchill} \&
  {Nielsen}}{{Kacprzak} et~al.}{2012}]{Kacprzak2012}
{Kacprzak} G.~G.,  {Churchill} C.~W.,    {Nielsen} N.~M.,  2012, \apjl, 760, L7

\bibitem[\protect\citeauthoryear{{Katz}, {Weinberg} \& {Hernquist}}{{Katz}
  et~al.}{1996}]{Katz1996}
{Katz} N.,  {Weinberg} D.~H.,    {Hernquist} L.,  1996, \apjs, 105, 19

\bibitem[\protect\citeauthoryear{{Katz} \& {White}}{{Katz} \&
  {White}}{1993}]{Katz1993}
{Katz} N.,  {White} S.~D.~M.,  1993, \apj, 412, 455

\bibitem[\protect\citeauthoryear{{Kere{\v s}}, {Katz}, {Fardal}, {Dav{\'e}} \&
  {Weinberg}}{{Kere{\v s}} et~al.}{2009}]{Keres2009}
{Kere{\v s}} D.,  {Katz} N.,  {Fardal} M.,  {Dav{\'e}} R.,    {Weinberg} D.~H.,
   2009, \mnras, 395, 160

\bibitem[\protect\citeauthoryear{{Kere{\v s}}, {Katz}, {Weinberg} \&
  {Dav{\'e}}}{{Kere{\v s}} et~al.}{2005}]{Keres2005}
{Kere{\v s}} D.,  {Katz} N.,  {Weinberg} D.~H.,    {Dav{\'e}} R.,  2005,
  \mnras, 363, 2

\bibitem[\protect\citeauthoryear{{Knollmann} \& {Knebe}}{{Knollmann} \&
  {Knebe}}{2009}]{Knollmann2009}
{Knollmann} S.~R.,  {Knebe} A.,  2009, \apjs, 182, 608

\bibitem[\protect\citeauthoryear{Kohler \& Gnedin}{Kohler \&
  Gnedin}{2007}]{Kohler2007}
Kohler K.,  Gnedin N.~Y.,  2007, ApJ, 655, 685

\bibitem[\protect\citeauthoryear{{Kollmeier}, {Miralda-Escud{\'e}}, {Cen} \&
  {Ostriker}}{{Kollmeier} et~al.}{2006}]{Kollmeier2006}
{Kollmeier} J.~A.,  {Miralda-Escud{\'e}} J.,  {Cen} R.,    {Ostriker} J.~P.,
  2006, \apj, 638, 52

\bibitem[\protect\citeauthoryear{{Lan}, {M{\'e}nard} \& {Zhu}}{{Lan}
  et~al.}{2014}]{Lan2014}
{Lan} T.-W.,  {M{\'e}nard} B.,    {Zhu} G.,  2014, \apj, 795, 31

\bibitem[\protect\citeauthoryear{Lehner, Howk, Tripp, Tumlinson, Prochaska,
  O'Meara, Thom, Werk, Fox \& Ribaudo}{Lehner et~al.}{2013}]{Lehner2013}
Lehner N.,  Howk J.~C.,  Tripp T.~M.,  Tumlinson J.,  Prochaska J.~X.,  O'Meara
  J.~M.,  Thom C.,  Werk J.~K.,  Fox a.~J.,    Ribaudo J.,  2013, ApJ, 770, 138

\bibitem[\protect\citeauthoryear{Lehner, O'Meara \& Fox}{Lehner
  et~al.}{2014}]{Lehner2014}
Lehner N.,  O'Meara J.,    Fox A.,  2014, ApJ, 788, 119

\bibitem[\protect\citeauthoryear{{Lehner}, {O'Meara}, {Howk}, {Prochaska} \&
  {Fumagalli}}{{Lehner} et~al.}{2016}]{2016arXiv160802588L}
{Lehner} N.,  {O'Meara} J.~M.,  {Howk} J.~C.,  {Prochaska} J.~X.,
  {Fumagalli} M.,  2016, arXiv:1608.02588

\bibitem[\protect\citeauthoryear{{Leitherer}, {Schaerer}, {Goldader},
  {Delgado}, {Robert}, {Kune}, {de Mello}, {Devost} \& {Heckman}}{{Leitherer}
  et~al.}{1999}]{Leitherer1999}
{Leitherer} C.,  {Schaerer} D.,  {Goldader} J.~D.,  {Delgado} R.~M.~G.,
  {Robert} C.,  {Kune} D.~F.,  {de Mello} D.~F.,  {Devost} D.,    {Heckman}
  T.~M.,  1999, \apjs, 123, 3

\bibitem[\protect\citeauthoryear{{Liang}, {Kravtsov} \& {Agertz}}{{Liang}
  et~al.}{2016}]{Liang2015}
{Liang} C.~J.,  {Kravtsov} A.~V.,    {Agertz} O.,  2016, \mnras, 458, 1164

\bibitem[\protect\citeauthoryear{{Ma}, {Hopkins}, {Faucher-Gigu{\`e}re},
  {Zolman}, {Muratov}, {Kere{\v s}} \& {Quataert}}{{Ma} et~al.}{2016}]{Ma2016}
{Ma} X.,  {Hopkins} P.~F.,  {Faucher-Gigu{\`e}re} C.-A.,  {Zolman} N.,
  {Muratov} A.~L.,  {Kere{\v s}} D.,    {Quataert} E.,  2016, \mnras, 456, 2140

\bibitem[\protect\citeauthoryear{McQuinn, Oh \& Faucher-Gigu{\`{e}}re}{McQuinn
  et~al.}{2011}]{McQuinn2011}
McQuinn M.,  Oh S.~P.,    Faucher-Gigu{\`{e}}re C.-A.,  2011, ApJ, 743, 82

\bibitem[\protect\citeauthoryear{{Martin}}{{Martin}}{2005}]{Martin2005}
{Martin} C.~L.,  2005, \apj, 621, 227

\bibitem[\protect\citeauthoryear{{Muratov}, {Keres}, {Faucher-Giguere},
  {Hopkins}, {Ma}, {Angles-Alcazar}, {Chan}, {Torrey}, {Hafen}, {Quataert} \&
  {Murray}}{{Muratov} et~al.}{2016}]{2016arXiv160609252M}
{Muratov} A.~L.,  {Keres} D.,  {Faucher-Giguere} C.-A.,  {Hopkins} P.~F.,  {Ma}
  X.,  {Angles-Alcazar} D.,  {Chan} T.~K.,  {Torrey} P.,  {Hafen} Z.~H.,
  {Quataert} E.,    {Murray} N.,  2016, arXiv:1606.09252

\bibitem[\protect\citeauthoryear{Muratov, Keres, Faucher-Giguere, Hopkins,
  Quataert \& Murray}{Muratov et~al.}{2015}]{Muratov2015}
Muratov A.~L.,  Keres D.,  Faucher-Giguere C.-A.,  Hopkins P.~F.,  Quataert E.,
     Murray N.,  2015, MNRAS, 454, 2691

\bibitem[\protect\citeauthoryear{{Murray}}{{Murray}}{2011}]{Murray2011}
{Murray} N.,  2011, \apj, 729, 133

\bibitem[\protect\citeauthoryear{{Oppenheimer} \& {Dav{\'e}}}{{Oppenheimer} \&
  {Dav{\'e}}}{2006}]{Oppenheimer2006}
{Oppenheimer} B.~D.,  {Dav{\'e}} R.,  2006, \mnras, 373, 1265

\bibitem[\protect\citeauthoryear{{Oppenheimer}, {Dav{\'e}}, {Kere{\v s}},
  {Fardal}, {Katz}, {Kollmeier} \& {Weinberg}}{{Oppenheimer}
  et~al.}{2010}]{Oppenheimer2010}
{Oppenheimer} B.~D.,  {Dav{\'e}} R.,  {Kere{\v s}} D.,  {Fardal} M.,  {Katz}
  N.,  {Kollmeier} J.~A.,    {Weinberg} D.~H.,  2010, \mnras, 406, 2325

\bibitem[\protect\citeauthoryear{{Peeples}, {Werk}, {Tumlinson}, {Oppenheimer},
  {Prochaska}, {Katz} \& {Weinberg}}{{Peeples} et~al.}{2014}]{Peeples2014}
{Peeples} M.~S.,  {Werk} J.~K.,  {Tumlinson} J.,  {Oppenheimer} B.~D.,
  {Prochaska} J.~X.,  {Katz} N.,    {Weinberg} D.~H.,  2014, \apj, 786, 54

\bibitem[\protect\citeauthoryear{{Planck Collaboration}, {Ade}, {Aghanim},
  {Arnaud}, {Ashdown}, {Aumont}, {Baccigalupi}, {Banday}, {Barreiro},
  {Bartlett} \& et al.}{{Planck Collaboration} et~al.}{2015}]{Planck2015}
{Planck Collaboration} {Ade} P.~A.~R.,  {Aghanim} N.,  {Arnaud} M.,  {Ashdown}
  M.,  {Aumont} J.,  {Baccigalupi} C.,  {Banday} A.~J.,  {Barreiro} R.~B.,
  {Bartlett} J.~G.,    et al. 2015, arXiv:1502.01589

\bibitem[\protect\citeauthoryear{{Porter}}{{Porter}}{1985}]{Porter1985}
{Porter} D.~H.,  1985, PhD thesis, California Univ., Berkeley.

\bibitem[\protect\citeauthoryear{{Price} \& {Monaghan}}{{Price} \&
  {Monaghan}}{2007}]{Price2007}
{Price} D.~J.,  {Monaghan} J.~J.,  2007, \mnras, 374, 1347

\bibitem[\protect\citeauthoryear{Prochaska \& Wolfe}{Prochaska \&
  Wolfe}{2009}]{Prochaska2009}
Prochaska J.~X.,  Wolfe A.~M.,  2009, ApJ, 696, 1543

\bibitem[\protect\citeauthoryear{{Prochaska et al.}}{{Prochaska et
  al.}}{2015}]{Prochaska2015}
{Prochaska et al.} 2015, \apjs, 221, 2

\bibitem[\protect\citeauthoryear{{Quiret}, {P{\'e}roux}, {Zafar}, {Kulkarni},
  {Jenkins}, {Milliard}, {Rahmani}, {Popping}, {Rao}, {Turnshek} \&
  {Monier}}{{Quiret} et~al.}{2016}]{2016MNRAS.458.4074Q}
{Quiret} S.,  {P{\'e}roux} C.,  {Zafar} T.,  {Kulkarni} V.~P.,  {Jenkins}
  E.~B.,  {Milliard} B.,  {Rahmani} H.,  {Popping} A.,  {Rao} S.~M.,
  {Turnshek} D.~A.,    {Monier} E.~M.,  2016, \mnras, 458, 4074

\bibitem[\protect\citeauthoryear{Rahmati, Pawlik, Rai{\v{c}}evic̀ \&
  Schaye}{Rahmati et~al.}{2013}]{Rahmati2013}
Rahmati A.,  Pawlik A.~H.,  Rai{\v{c}}evic̀ M.,    Schaye J.,  2013, MNRAS,
  430, 2427

\bibitem[\protect\citeauthoryear{{Rahmati}, {Schaye}, {Bower}, {Crain},
  {Furlong}, {Schaller} \& {Theuns}}{{Rahmati} et~al.}{2015}]{Rahmati2015a}
{Rahmati} A.,  {Schaye} J.,  {Bower} R.~G.,  {Crain} R.~A.,  {Furlong} M.,
  {Schaller} M.,    {Theuns} T.,  2015, \mnras, 452, 2034

\bibitem[\protect\citeauthoryear{Ribaudo, Lehner \& Howk}{Ribaudo
  et~al.}{2011}]{Ribaudo2011}
Ribaudo J.,  Lehner N.,    Howk J.~C.,  2011, ApJ, 736, 42

\bibitem[\protect\citeauthoryear{{Ribaudo}, {Lehner}, {Howk}, {Werk}, {Tripp},
  {Prochaska}, {Meiring} \& {Tumlinson}}{{Ribaudo} et~al.}{2011}]{Ribaudo2011a}
{Ribaudo} J.,  {Lehner} N.,  {Howk} J.~C.,  {Werk} J.~K.,  {Tripp} T.~M.,
  {Prochaska} J.~X.,  {Meiring} J.~D.,    {Tumlinson} J.,  2011, \apj, 743, 207

\bibitem[\protect\citeauthoryear{{Rubin}, {Prochaska}, {Koo}, {Phillips},
  {Martin} \& {Winstrom}}{{Rubin} et~al.}{2014}]{Rubin2014}
{Rubin} K.~H.~R.,  {Prochaska} J.~X.,  {Koo} D.~C.,  {Phillips} A.~C.,
  {Martin} C.~L.,    {Winstrom} L.~O.,  2014, \apj, 794, 156

\bibitem[\protect\citeauthoryear{{Schaye}, {Carswell} \& {Kim}}{{Schaye}
  et~al.}{2007}]{2007MNRAS.379.1169S}
{Schaye} J.,  {Carswell} R.~F.,    {Kim} T.-S.,  2007, \mnras, 379, 1169

\bibitem[\protect\citeauthoryear{{Shapley}, {Steidel}, {Pettini} \&
  {Adelberger}}{{Shapley} et~al.}{2003}]{Shapley2003}
{Shapley} A.~E.,  {Steidel} C.~C.,  {Pettini} M.,    {Adelberger} K.~L.,  2003,
  \apj, 588, 65

\bibitem[\protect\citeauthoryear{{Shen}, {Madau}, {Aguirre}, {Guedes}, {Mayer}
  \& {Wadsley}}{{Shen} et~al.}{2012}]{2012ApJ...760...50S}
{Shen} S.,  {Madau} P.,  {Aguirre} A.,  {Guedes} J.,  {Mayer} L.,    {Wadsley}
  J.,  2012, \apj, 760, 50

\bibitem[\protect\citeauthoryear{{Shen}, {Wadsley} \& {Stinson}}{{Shen}
  et~al.}{2010}]{2010MNRAS.407.1581S}
{Shen} S.,  {Wadsley} J.,    {Stinson} G.,  2010, \mnras, 407, 1581

\bibitem[\protect\citeauthoryear{Somerville \& Dav{\'{e}}}{Somerville \&
  Dav{\'{e}}}{2015}]{Somerville2015}
Somerville R.~S.,  Dav{\'{e}} R.,  2015, Annu. Rev. Astron. Astrophys., 53, 51

\bibitem[\protect\citeauthoryear{Sparre, Hayward, Feldmann,
  Faucher-Gigu{\`{e}}re, Muratov, Kere{\v{s}} \& Hopkins}{Sparre
  et~al.}{2015}]{Sparre2015}
Sparre M.,  Hayward C.~C.,  Feldmann R.,  Faucher-Gigu{\`{e}}re C.-A.,  Muratov
  A.~L.,  Kere{\v{s}} D.,    Hopkins P.~F.,  2015, arXiv:1510.03869

\bibitem[\protect\citeauthoryear{Springel}{Springel}{2005}]{Springel2005}
Springel V.,  2005, MNRAS, 364, 1105

\bibitem[\protect\citeauthoryear{Steidel, Erb, Shapley, Pettini, Reddy,
  Bogosavljevi{\'{c}}, Rudie \& Rakic}{Steidel et~al.}{2010}]{Steidel2010}
Steidel C.~C.,  Erb D.~K.,  Shapley A.~E.,  Pettini M.,  Reddy N.,
  Bogosavljevi{\'{c}} M.,  Rudie G.~C.,    Rakic O.,  2010, ApJ, 717, 289

\bibitem[\protect\citeauthoryear{{Tremonti}, {Heckman}, {Kauffmann},
  {Brinchmann}, {Charlot}, {White}, {Seibert}, {Peng}, {Schlegel}, {Uomoto},
  {Fukugita} \& {Brinkmann}}{{Tremonti} et~al.}{2004}]{Tremonti2004}
{Tremonti} C.~A.,  {Heckman} T.~M.,  {Kauffmann} G.,  {Brinchmann} J.,
  {Charlot} S.,  {White} S.~D.~M.,  {Seibert} M.,  {Peng} E.~W.,  {Schlegel}
  D.~J.,  {Uomoto} A.,  {Fukugita} M.,    {Brinkmann} J.,  2004, \apj, 613, 898

\bibitem[\protect\citeauthoryear{{Tripp}, {Meiring}, {Prochaska}, {Willmer},
  {Howk}, {Werk}, {Jenkins}, {Bowen}, {Lehner}, {Sembach}, {Thom} \&
  {Tumlinson}}{{Tripp} et~al.}{2011}]{Tripp2011}
{Tripp} T.~M.,  {Meiring} J.~D.,  {Prochaska} J.~X.,  {Willmer} C.~N.~A.,
  {Howk} J.~C.,  {Werk} J.~K.,  {Jenkins} E.~B.,  {Bowen} D.~V.,  {Lehner} N.,
  {Sembach} K.~R.,  {Thom} C.,    {Tumlinson} J.,  2011, Science, 334, 952

\bibitem[\protect\citeauthoryear{{van de Voort}, {Schaye}, {Altay} \&
  {Theuns}}{{van de Voort} et~al.}{2012}]{vdVoort12}
{van de Voort} F.,  {Schaye} J.,  {Altay} G.,    {Theuns} T.,  2012, \mnras,
  421, 2809

\bibitem[\protect\citeauthoryear{{van de Voort}, {Schaye}, {Booth}, {Haas} \&
  {Dalla Vecchia}}{{van de Voort} et~al.}{2011}]{vdVoort2011}
{van de Voort} F.,  {Schaye} J.,  {Booth} C.~M.,  {Haas} M.~R.,    {Dalla
  Vecchia} C.,  2011, \mnras, 414, 2458

\bibitem[\protect\citeauthoryear{Watson, Iliev, D'Aloisio, Knebe, Shapiro \&
  Yepes}{Watson et~al.}{2013}]{Watson2013a}
Watson W.~A.,  Iliev I.~T.,  D'Aloisio A.,  Knebe A.,  Shapiro P.~R.,    Yepes
  G.,  2013, MNRAS, 433, 1230

\bibitem[\protect\citeauthoryear{{Weiner}, {Coil}, {Prochaska}, {Newman},
  {Cooper}, {Bundy}, {Conselice}, {Dutton}, {Faber}, {Koo}, {Lotz}, {Rieke} \&
  {Rubin}}{{Weiner} et~al.}{2009}]{Weiner2009}
{Weiner} B.~J.,  {Coil} A.~L.,  {Prochaska} J.~X.,  {Newman} J.~A.,  {Cooper}
  M.~C.,  {Bundy} K.,  {Conselice} C.~J.,  {Dutton} A.~A.,  {Faber} S.~M.,
  {Koo} D.~C.,  {Lotz} J.~M.,  {Rieke} G.~H.,    {Rubin} K.~H.~R.,  2009, \apj,
  692, 187

\bibitem[\protect\citeauthoryear{{Wotta}, {Lehner}, {Howk}, {O'Meara} \&
  {Prochaska}}{{Wotta} et~al.}{2016}]{2016ApJ...831...95W}
{Wotta} C.~B.,  {Lehner} N.,  {Howk} J.~C.,  {O'Meara} J.~M.,    {Prochaska}
  J.~X.,  2016, \apj, 831, 95

\bibitem[\protect\citeauthoryear{{Zahid}, {Kewley} \& {Bresolin}}{{Zahid}
  et~al.}{2011}]{Zahid2011}
{Zahid} H.~J.,  {Kewley} L.~J.,    {Bresolin} F.,  2011, \apj, 730, 137

\end{thebibliography}
 
\end{document}